\documentclass[twocolumn]{aastex61}
\pdfoutput=1 
\usepackage[utf8]{inputenc}
\usepackage{hyperref}
\usepackage{natbib}
\usepackage{graphicx}
\usepackage{xspace}
\usepackage[caption=false]{subfig}

\newcommand{\rearth}{$R_\earth$\xspace}
\newcommand{\rearthe}{$R_\earth$}

\newcommand{\rjup}{$R_{\rm Jup}~$}
\newcommand{\rsun}{\ensuremath{R_{\odot}}\xspace}
\newcommand{\mstar}{\ensuremath{M_{\star}}\xspace}
\newcommand{\rstar}{\ensuremath{R_{\star}}\xspace}

\newcommand{\feh}{\ensuremath{[\mbox{Fe}/\mbox{H}]}\xspace}
\newcommand{\teff}{\ensuremath{T_{\mathrm{eff}}}\xspace}
\newcommand{\logg}{\ensuremath{\log g}\xspace}

\newcommand{\Rp}{$R_p$\xspace}

\newcommand{\kepler}{\textit{Kepler}\xspace}
\newcommand{\ktwo}{\textit{K2}\xspace}
\newcommand{\spitzer}{\textit{Spitzer}\xspace}
\newcommand{\tess}{\textit{TESS}\xspace}
\newcommand{\jwst}{\textit{JWST}\xspace}
\newcommand{\cheops}{\textit{CHEOPS}\xspace}
\newcommand{\wise}{\textit{WISE}\xspace}
\newcommand{\ktwophot}{\texttt{k2phot}\xspace}
\newcommand{\terra}{\texttt{TERRA}\xspace}

\newcommand{\vespa}{\texttt{vespa}\xspace}

\newcommand{\nsmall}{\ensuremath{18}\xspace}
\newcommand{\nbright}{five\xspace}
\newcommand{\ntotal}{\ensuremath{155}\xspace}
\newcommand{\nvalidated}{\ensuremath{60}\xspace}
\newcommand{\ncandidates}{\ensuremath{77}\xspace}
\newcommand{\nfp}{18\xspace}
\newcommand{\nalreadyvalid}{\ensuremath{40}\xspace}
\newcommand{\nnewlyvalid}{\ensuremath{20}\xspace}
\newcommand{\nmultip}{\ensuremath{24}\xspace}
\newcommand{\nmultis}{11\xspace}
\newcommand{\nusp}{three\xspace}
\newcommand{\medrad}{\ensuremath{2.5}\,\rearthe \xspace}
\newcommand{\medper}{\ensuremath{7.1}\,d\xspace}
\newcommand{\medteq}{\ensuremath{811}\,K\xspace}
\newcommand{\medjmag}{\ensuremath{11.3}\,mag\xspace}
\newcommand{\minfpp}{1\%\xspace}
\newcommand{\maxfpp}{99\%\xspace}
\newcommand{\nexsciquerydate}{August 15, 2018}

\shorttitle{{\it K2} Year 2}
\shortauthors{Livingston et al.}

\begin{document}

\title{\nvalidated validated planets from \ktwo campaigns 5--8}

\author{John H. Livingston}
\affiliation{Department of Astronomy, University of Tokyo, 7-3-1 Hongo, Bunkyo-ku, Tokyo 113-0033, Japan}
\affiliation{JSPS Fellow}
\affiliation{\href{mailto:livingston@astron.s.u-tokyo.ac.jp}{{\tt livingston@astron.s.u-tokyo.ac.jp}}}

\author{Ian J. M. Crossfield}
\affiliation{Department of Physics, Massachusetts Institute of Technology, Cambridge, MA 02139, USA}

\author{Erik A. Petigura}
\affiliation{Geological and Planetary Sciences, California Institute of Technology, 1200 E California Blvd, Pasadena, CA 91125, USA}
\affiliation{Hubble Fellow}

\author{Erica J. Gonzales}
\affiliation{Department of Astronomy, University of California, Santa Cruz, Santa Cruz, CA, USA}
\affiliation{NSF GRFP Fellow}

\author{David R. Ciardi}
\affiliation{Caltech/IPAC-NASA Exoplanet Science Institute, M/S 100-22, 770 S. Wilson Ave, Pasadena, CA 91106 USA}

\author{Charles A. Beichman}
\affiliation{Caltech/IPAC-NASA Exoplanet Science Institute, M/S 100-22, 770 S. Wilson Ave, Pasadena, CA 91106 USA}

\author{Jessie L. Christiansen}
\affiliation{Caltech/IPAC-NASA Exoplanet Science Institute, M/S 100-22, 770 S. Wilson Ave, Pasadena, CA 91106 USA}

\author{Courtney D. Dressing}
\affiliation{Department of Astronomy, University of California at Berkeley, Berkeley, CA 94720, USA}

\author{Thomas Henning}
\affiliation{Max Planck Institut f\"ur Astronomie, K\"onigstuhl 17, 69117 Heidelberg, Germany}

\author{Andrew W. Howard}
\affiliation{Department of Astronomy, California Institute of Technology, Pasadena, CA 91125, USA}

\author{Howard Isaacson}
\affiliation{Department of Astronomy, University of California at Berkeley, Berkeley, CA 94720, USA}

\author{Benjamin J. Fulton}
\affiliation{Department of Astronomy, California Institute of Technology, Pasadena, CA 91125, USA}

\author{Molly Kosiarek}
\affiliation{Department of Astronomy, University of California, Santa Cruz, Santa Cruz, CA, USA}

\author{Joshua E. Schlieder}
\affiliation{Exoplanets and Stellar Astrophysics Laboratory, Code 667, NASA Goddard Space Flight Center, Greenbelt, MD 20771, USA}

\author{Evan Sinukoff}
\affiliation{Department of Astronomy, California Institute of Technology, Pasadena, CA 91125, USA}
\affiliation{Institute for Astronomy, University of Hawai`i at M\={a}noa, Honolulu, HI, USA}

\author{Motohide Tamura}
\affiliation{Department of Astronomy, University of Tokyo, 7-3-1 Hongo, Bunkyo-ku, Tokyo 113-0033, Japan}
\affiliation{Astrobiology Center, NINS, 2-21-1 Osawa, Mitaka, Tokyo 181-8588, Japan}
\affiliation{National Astronomical Observatory of Japan, NINS, 2-21-1 Osawa, Mitaka, Tokyo 181-8588, Japan}

\begin{abstract}

We present a uniform analysis of \ntotal candidates from the second year of NASA's \ktwo mission (Campaigns 5--8), yielding \nvalidated statistically validated planets spanning a range of properties, with median values of \Rp = \medrad, $P$ = \medper, $T_\mathrm{eq}$ = \medteq, and {\em J} = \medjmag. The sample includes \nmultip planets in \nmultis multi-planetary systems, as well as \nfp false positives, and \ncandidates remaining planet candidates. Of particular interest are \nsmall planets smaller than 2\,\rearth, \nbright orbiting stars brighter than $J$\,=\,10\,mag, and a system of four small planets orbiting the solar-type star EPIC\,212157262. We compute planetary transit parameters and false positive probabilities using a robust statistical framework and present a complete analysis incorporating the results of an intensive campaign of high resolution imaging and spectroscopic observations. This work brings the \ktwo yield to over 360 planets, and by extrapolation we expect that \ktwo will have discovered $\sim$600 planets before the expected depletion of its on-board fuel in late 2018.

\end{abstract}

\section{Introduction}
%

The \kepler mission provided a trove of data unprecedented in both quality and quantity, which opened new vistas to planet occurrence and diversity. In addition to revolutionizing the study of planetary demographics through the discovery of over 2000 validated planets, \kepler has enabled us to address questions about the abundance of Earth-sized worlds \citep{2013PNAS..11019273P, 2013ApJ...767...95D, 2014ApJ...795...64F, 2015ApJ...807...45D, 2015ApJ...809....8B}. However, after the mechanical failure of a second reaction wheel, the spacecraft was no longer able to point with the stability required for its prime mission, which led to the new mode of operation known as the \ktwo mission \citep{2014PASP..126..398H}.

\ktwo continues the legacy of \kepler by discovering large numbers of planets, while pursuing a wider and shallower survey than the original \kepler mission. To date, \ktwo has significantly enhanced the number of known planets orbiting bright and/or late-type host stars, as compared to those stars surveyed by \kepler \citep[e.g.][]{2015ApJ...806..215F, 2015ApJ...809...25M, 2016ApJS..222...14V, 2016ApJS..226....7C, 2016MNRAS.461.3399P, 2017AJ....154..207D, 2018AJ....155..127H}. It has also discovered planets in cluster environments \citep[e.g.][]{2016AJ....152..223O, 2016AJ....151..112D, 2017MNRAS.464..850G, 2018AJ....155...10C}, including a 5--10 Myr planet in the Upper Scorpius star forming region \citep{2016Natur.534..658D, 2016AJ....152...61M}.

By observing a succession of fields along the ecliptic plane, \ktwo compensates for the decreased pointing stability of the \kepler spacecraft by minimizing torque from solar radiation pressure. \ktwo's wide survey and community-led target selection has enabled it to observe a greater number of nearby stars as well as probe a greater diversity of stellar environments. This has led to the discovery of many planets orbiting bright stars which are more suitable to follow-up studies than those found by \kepler. Space-based transit surveys like \kepler are efficient methods to find candidate planets. However, such surveys are also efficient at finding false positives, namely diluted eclipsing binaries. Cleaning these samples of false positives is crucial for demographic work as well as for efficient utilization of follow-up resources.

While some planets can be confirmed based on direct detection of stellar reflex motion \citep[RVs; e.g.][]{1952Obs....72..199S, 1995Natur.378..355M} or planet-planet gravitation interactions \citep[TTVs; e.g.][]{2005Sci...307.1288H, 2005MNRAS.359..567A}, this is not possible for the vast majority of \kepler and \ktwo candidates, as the host stars are too numerous and generally too faint for RV confirmation {\it en masse}, and TTVs are only detectable for a subset of multi-planet systems. Instead, we turn to statistical validation, where we calculate each planet candidate's false positive probability \citep[FPP; e.g.][]{2011ApJ...727...24T, 2012ApJ...761....6M, 2014MNRAS.441..983D}. This approach yields the probability that a candidates is a real planet given the light curve shape, stellar properties, and constraints on nearby companions, and has been used to validate thousands of new planets \citep{2016ApJ...822...86M}.

In this paper, we apply the tools of statistical planet validation to \ntotal candidates detected by our team using data from \ktwo's second year of operations (Campaigns 5--8), incorporating stellar characterization and limits on close companions from high resolution imaging. This work builds off of the work presented in \citet{2016ApJS..226....7C}. Our analysis incorporates the results of several companion papers: \citet[][hereafter P18]{2018AJ....155...21P} describes our photometry and transit search pipeline and presents a catalog of vetted candidates for C5--8 along with stellar spectroscopy for FGK host stars; Gonzales et al. (2018, in preparation, hereafter G18) presents our high resolution imaging, with which we detect faint nearby companions and produce contrast curves used in the validation process. Because statistical validation depends on the presence of nearby companions, stellar properties, and light curve shape parameters, we synthesize planet and host star properties from our various analyses in order to compute valid FPPs. The result is a catalog of new planetary systems, some of which are interesting targets for future studies. In particular, Doppler mass measurements with high precision spectrographs will enable better understanding of bulk planet composition and formation/migration histories, and transmission/emission spectroscopy with \jwst will probe previously unexplored atmospheric regimes.

The structure of this paper is as follows. In \autoref{sec:candidates} we provide an overview of our \ktwo target selection, photometry, and transit search, which results in the set of planet candidates for which we conduct follow-up observations and validation. \autoref{sec:hosts} and \autoref{sec:lightcurves} describe our host star characterization and light curve analyses, respectively. In \autoref{sec:validation} we describe our validation procedures, and in \autoref{sec:discussion} we discuss the overall results as well as particular systems of interest, concluding with a summary in \autoref{sec:summary}.

\section{Identification of planet candidates}
\label{sec:candidates}

\begin{figure*}
    \centering
    \includegraphics[width=0.84\textwidth,trim={0cm 0cm 0cm 0cm}]{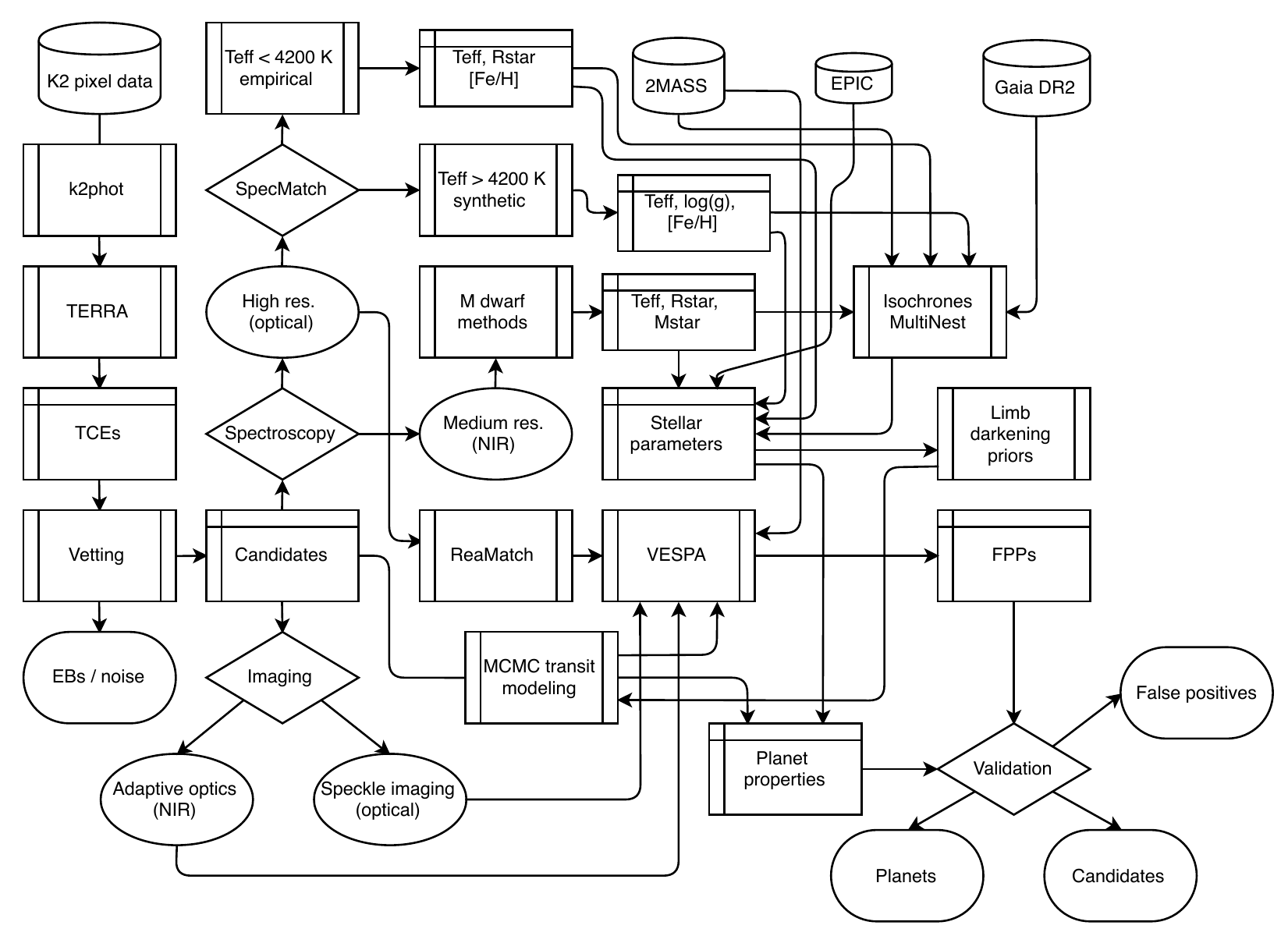}
    \caption{A schematic overview of the validation process, beginning with the \ktwo pixel data, ending with candidate dispositions, and including various follow-up observations as well as intermediate analyses. Cylinders represent external data sources, rectangles represent specific processes/codes (left/right sides double-lined) and their data products (left/top double-lined), diamonds represent general observations/analyses, ovals represent specific analyses, and round-edged rectangles represent the final validation dispositions.}
    \label{fig:flowchart}
\end{figure*}

\subsection{Target selection}

Our team successfully proposed \ktwo General Observer (GO) targets for Campaigns 5--8.\footnote{GO Programs 5011, 5033, 5046, 6008, 6030, 7008, 7030, 7043, 8012, 8056, and 8077} In brief, we used data from the \tess Dwarf Catalog \citep{2014arXiv1410.6379S}, the SUPERBLINK proper motion database \citep{2005AJ....129.1483L}, the PanSTARRS-1 survey \citep{2002SPIE.4836..154K,2016arXiv161205560C}, 2MASS, and \wise, applying color and proper motion cuts in order to select solar- and late-type dwarf stars, while minimizing contamination from background giants (for a more detailed description, see \citealt{2016ApJS..226....7C} and P18). As the \ktwo data from all GO programs are public, we have included data besides those from our own proposals in our search for candidate planet transit signals. For reference, we have listed all GO programs associated with each of the targets in this work in \autoref{tab:stellar}.

\subsection{Photometry and transit search}

\begin{figure*}
    \centering
    \includegraphics[width=0.84\textwidth,trim={0cm 0.5cm 0cm 0}]{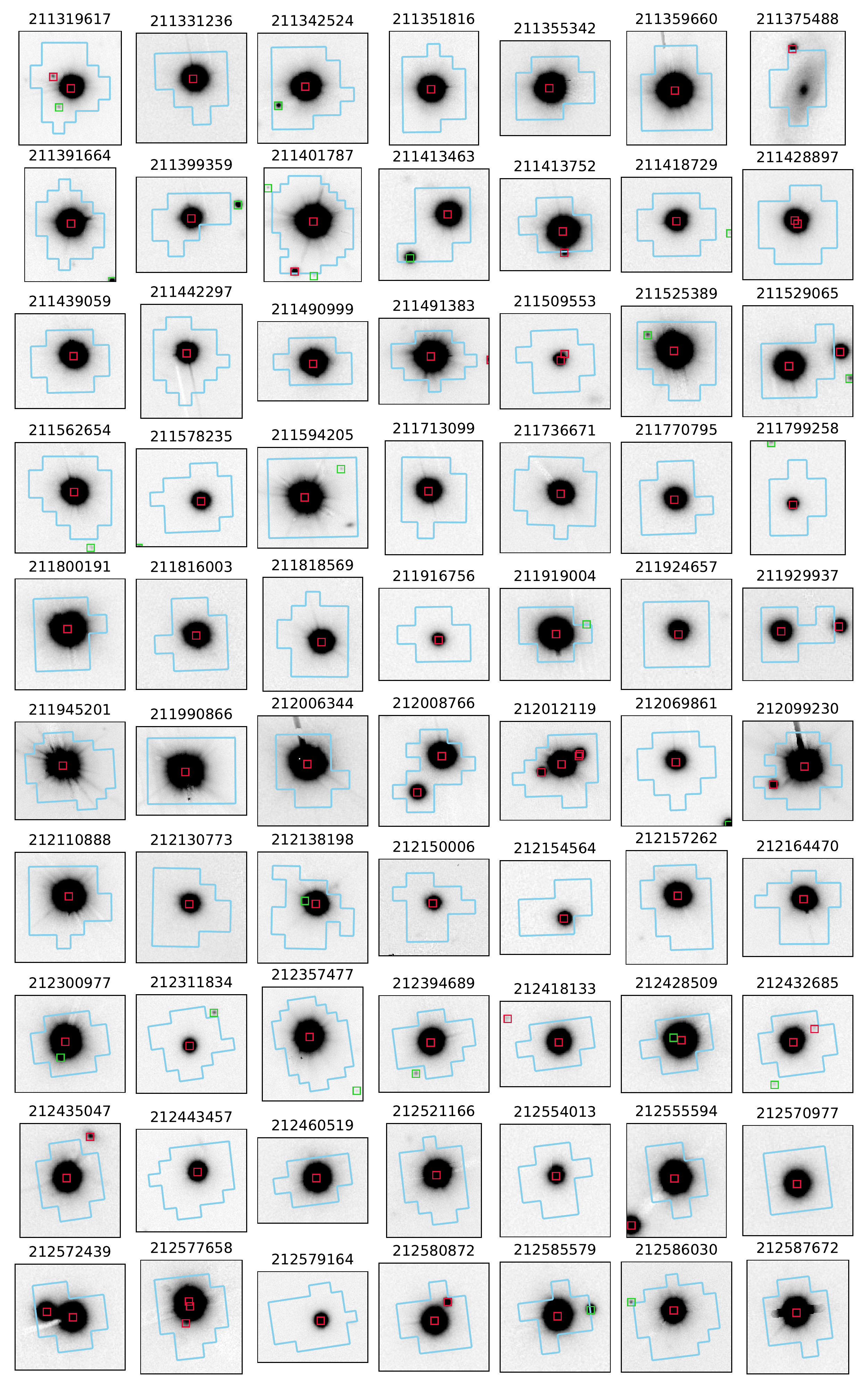}
    \caption{PanSTARRS-1 $r$-band images with {\tt k2phot} optimal apertures overplotted in blue, which yield the light curves analyzed in this work (except for the candidate 211978865.01). {\it Gaia} DR2 sources (open squares) are colored as described in \autoref{sec:companions}.}
    \label{fig:aper1}
\end{figure*}

\begin{figure*}
    \centering
    \includegraphics[width=0.84\textwidth,trim={0cm 0.5cm 0cm 0}]{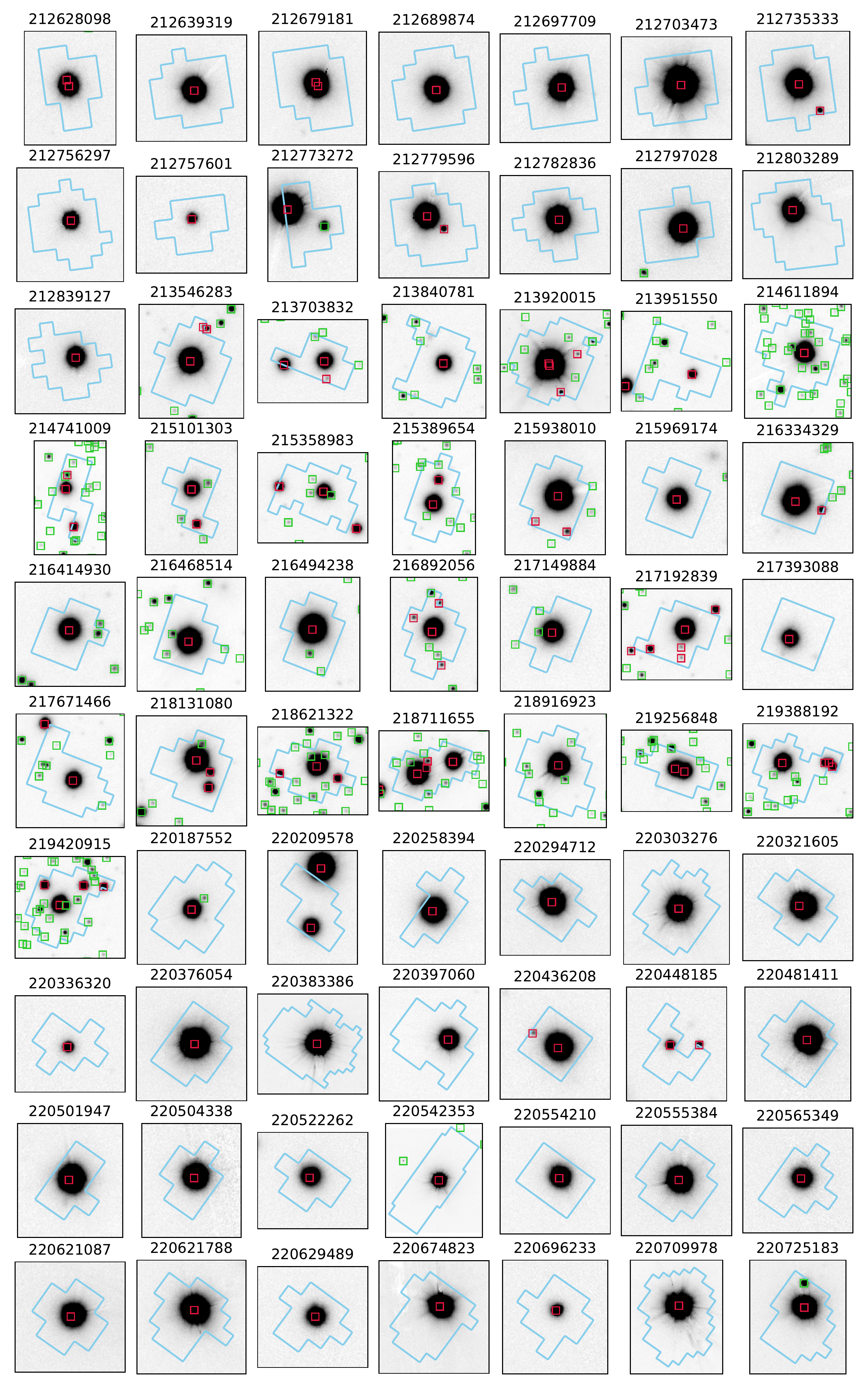}
    \caption{Continuation of \autoref{fig:aper1}.}
    \label{fig:aper2}
\end{figure*}

As described in P18, we used the publicly available software packages \ktwophot\footnote{\url{https://github.com/petigura/k2phot}} \citep{2015ApJ...811..102P} and \terra\footnote{\url{https://github.com/petigura/terra}} to produce calibrated photometric time-series from the \ktwo pixel data for 87,913 stars from C5--8 and identify planet candidates. In brief, each calibrated light curve is iteratively searched for transit-like signals, by masking the transits of each successive candidate identified and repeating the search. This iterative approach allows us to detect multi-candidate systems. These ``threshold-crossing events'' (TCEs) are then subjected to further scrutiny in order to identify obviously spurious signals and minimize the number of false positives in our candidate sample.
\autoref{fig:flowchart} presents an overview of how our photometry and transit search fits into the process of candidate identification, follow-up observations, and detailed analyses. For a full description of our photometry, transit search, and candidate vetting procedures, see \citet{2016ApJS..226....7C} and P18. In addition to the 151 planet candidates reported by P18, we identified four candidates in the light curves of stars already reported to have at least one candidate by P18. The analysis that follows considers the resulting set of \ntotal planet candidates orbiting the same set of 141 stars as analyzed by P18. \autoref{fig:aper1} and \autoref{fig:aper2} show 1\arcmin$\times$1\arcmin\, $r$-band image stamps from PanSTARRS-1 with {\tt k2phot} optimal apertures overplotted for the stars we analyze here.

\section{Host characterization}
\label{sec:hosts}

\subsection{High resolution imaging and companion search}
\label{sec:imaging}

From 2016 January 26 to 2017 August 20 UT, we performed high resolution imaging follow-up observations to identify stellar companions. We employed adaptive optics (AO) techniques using the following near-infrared (NIR) cameras: NIRC2 \citep{2014SPIE.9148E..2BW} on the 10-m Keck II telescope, PHARO \citep{2001PASP..113..105H} on the 5-m Hale telescope, and NIRI on the 8-m Gemini North \citep{2003PASP..115.1388H}. For all instruments, initial detection of diluting companions is conducted by observing in the $K$ band (centered at 2.196 $\mu$m), $K_\mathrm{cont}$ (centered at 2.27 $\mu$m), or Br-$\gamma$ (centered at 2.168 $\mu$m). Some targets were also observed in $J$ (centered at 1.248 $\mu$m) in order to obtain NIR colors of any detected secondary sources. Efforts are currently underway to obtain multi-band observations of targets with diluting companions, so as to ascertain the bound or unbound nature of the companion. For further details of the NIR AO imaging follow-up see G18.

Speckle-interferometric observations were also conducted in the optical for most targets using the Differential Speckle Survey Instrument (DSSI) \citep{2009AJ....137.5057H,2012AJ....144..165H} on the Gemini 8-m telescopes, and the NN-EXPLORE Exoplanet and Stellar Speckle Imager (NESSI) \citep{2011AJ....142...19H, 2016SPIE.9907E..2RS} on the WIYN 3.5-m telescope. For further details of the optical speckle-interferometric follow-up, see \citet{2018AJ....156...31M}.

\begin{figure}
    \centering
    \includegraphics[width=0.48\textwidth,trim={0.25cm 0cm 0.0cm 0}]{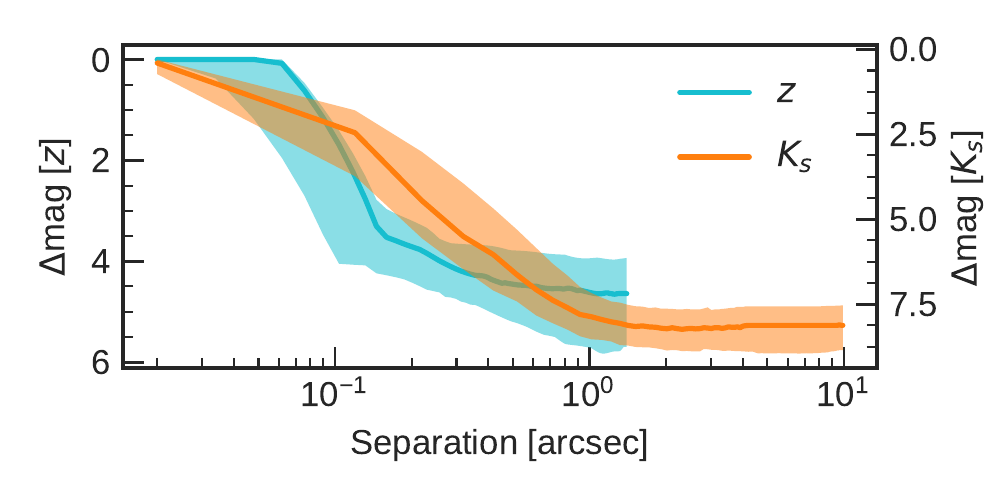}
    \caption{The median and 16$^\mathrm{th}$ -- 84$^\mathrm{th}$ percentile range of the $z$- and $K_s$-band contrast curves used in this work. The $z$-band contrast curves are derived from speckle imaging observations, which have a smaller field of view than the $K_s$-band AO images.}
    \label{fig:cc}
\end{figure}

The contrast curves derived from these high resolution images are an important constraint in the calculation of statistical FPPs, as they place limits on the existence of nearby bound stellar companions or background stars which could be the source of the observed transit signals. To illustrate the typical strength of these constraints, we compute the median $z$- and $K_s$-band contrast curves used in this work, derived from speckle imaging and AO, respectively. We plot these median contrast curves along with their 16$^\mathrm{th}$ -- 84$^\mathrm{th}$ percentile ranges in \autoref{fig:cc}.

\subsection{Spectroscopy and stellar parameters}
\label{sec:stellar}
\begin{figure}
    \centering
    \includegraphics[width=0.45\textwidth,trim={0cm 0cm 0cm 0}]{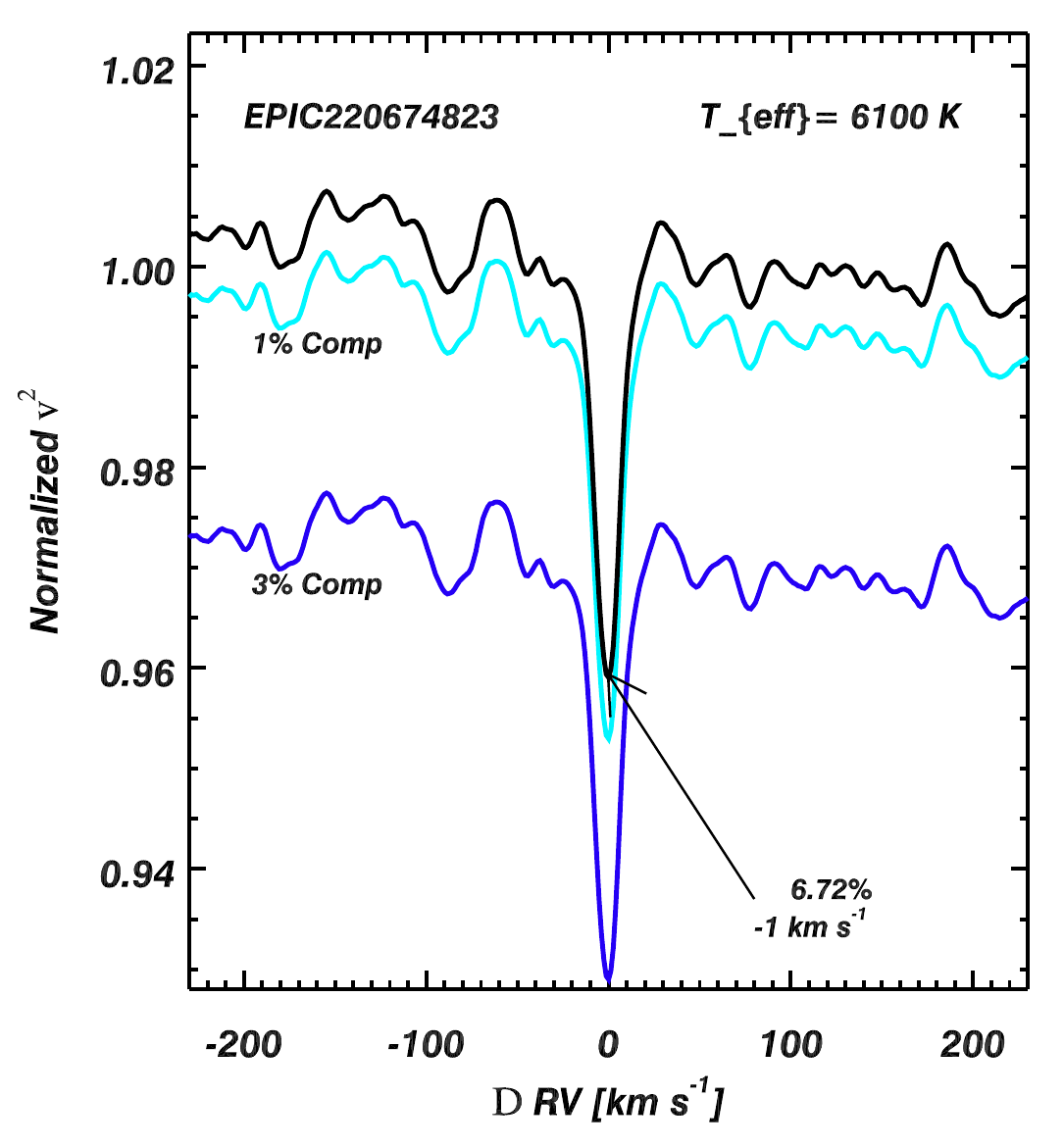}
    \caption{An example of the spectroscopic binary search analysis for K2-106 performed by {\tt ReaMatch} \citep{2015AJ....149...18K}.}
    \label{fig:reamatch}
\end{figure}

We obtained high resolution optical spectra for most of the targets in this work using Keck/HIRES, described in detail in P18. From these spectra we derive stellar parameters using {\tt SpecMatch-syn} \citep{2017AJ....154..107P} for stars hotter than 4200 K, and {\tt SpecMatch-emp} \citep{2017ApJ...836...77Y} for cooler stars. In addition, we refer the reader to \citet{2017ApJ...836..167D} and \citet{2017ApJ...837...72M}, in which spectroscopic analyses of many of the M dwarfs in the sample were presented. As an input to \vespa, we adopt the constraints on secondary stars determined by {\tt ReaMatch} \citep{2015AJ....149...18K} (presented in P18), which are typically $\Delta V \leq 5$ mag for $v\sin i \geq 10$ km s$^{-1}$. We show a plot illustrating this analysis in \autoref{fig:reamatch}. For a subset of 21 late-type stars in this work, we adopt the stellar parameters of \citet{2017ApJ...836..167D} and \citet{2017ApJ...837...72M}, who obtained medium resolution NIR spectra with IRTF/SpeX and NTT/SOFI. In total, 119 of the host stars we analyze here have spectroscopically-derived parameters.

To facilitate a uniform analysis of all the candidates, we utilized the Python package {\tt isochrones} to infer stellar parameters using priors from the aforementioned spectroscopic analyses, 2MASS $JHK$ photometry \citep{2006AJ....131.1163S}, and {\it Gaia} DR2 parallaxes \citep{2016A&A...595A...1G, 2018A&A...616A...1G}. This step is important because we have a heterogeneous set of parameters derived from spectroscopy, and 22 stars lack spectra entirely. From {\tt SpecMatch-syn}, we have the parameters effective temperature \teff, surface gravity \logg, and metallicity [Fe/H], whereas {\tt SpecMatch-emp} yields \teff, [Fe/H], and radius \rstar. The NIR spectroscopic analyses \citet{2017ApJ...836..167D} and \citet{2017ApJ...837...72M} are based on empirical relations calibrated to nearby stars with interferometrically-measured radii, and yield the parameters of interest \teff, \rstar, and mass \mstar. For each star, we estimated the missing parameters using {\tt isochrones}, which uses MultiNest \citep{2013arXiv1306.2144F} in conjunction with the Dartmouth stellar evolution models \citep{2008ApJS..178...89D}, effectively combining prior knowledge from spectroscopy, photometry, and parallax to constrain all parameters of interest. The resulting stellar parameters used in this work are listed in \autoref{tab:stellar}.

\begin{figure*}
    \centering
    \includegraphics[width=0.99\textwidth,trim={0.5cm 0 0.5cm 0}]{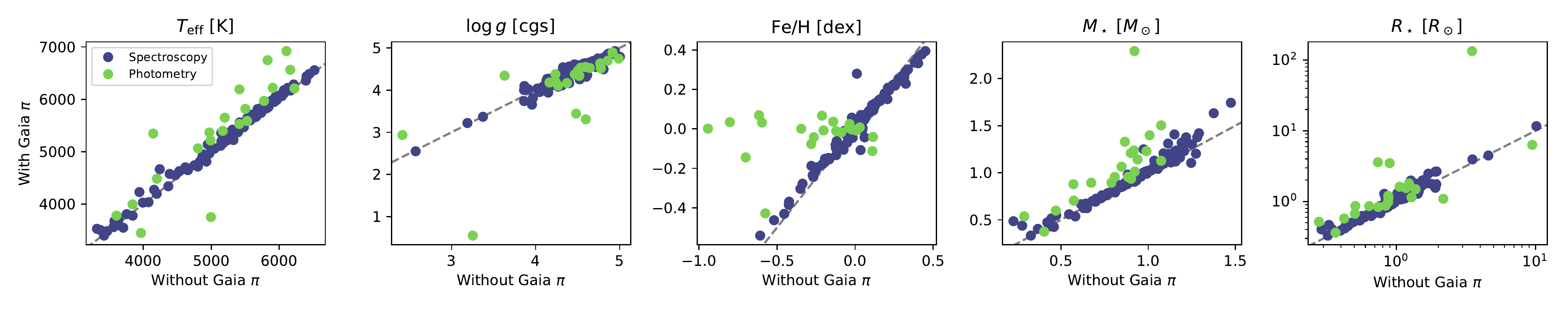}
    \caption{Visualization of the stellar parameters used in this work. The blue and green points correspond to stars with and without high resolution optical or medium resolution NIR spectroscopic constraints, respectively (see \autoref{sec:stellar}). The $x$- and $y$-axes correspond to the value of each stellar parameter before and after incorporating parallaxes from {\it Gaia} DR2, respectively, and the dashed gray lines indicate equality.}
    \label{fig:stellar}
\end{figure*}

For the 22 stars in this sample that lack spectroscopic constraints, we compare the parameters we derive from {\it Gaia} DR2 parallax and 2MASS $JHK$ photometry to the parameters from the Ecliptic Plane Input Catalog \citep[EPIC;][]{2016ApJS..224....2H}, which were based on photometry, proper motion, and models of the distribution of stars in the milky way. We plot these parameters in \autoref{fig:stellar}, highlighting the significance of parallax for stars lacking spectroscopy. The additional constraint from parallax reveals that several of these are late-type stars with underestimated radii in the EPIC. \citet{2016ApJS..224....2H} were aware of this systematic effect, which was the result of their choice of isochrones; this bias was later empirically shown to be $\sim$40\% for M dwarfs by \citet{2017ApJ...836..167D}. Including parallax also eliminates most of the potential for misclassifying dwarfs and sub-giants, a frequent problem with the parameters in the EPIC. However, spectroscopic observations would be useful to more precisely constrain the radii of planets orbiting these stars.

\section{Light curve analyses}
\label{sec:lightcurves}

\subsection{Transit modeling}
\label{sec:modeling}

\begin{figure*}
    \centering
    \includegraphics[width=0.95\textwidth,trim={0.5cm 0.5cm 0.5cm 0.5cm}]{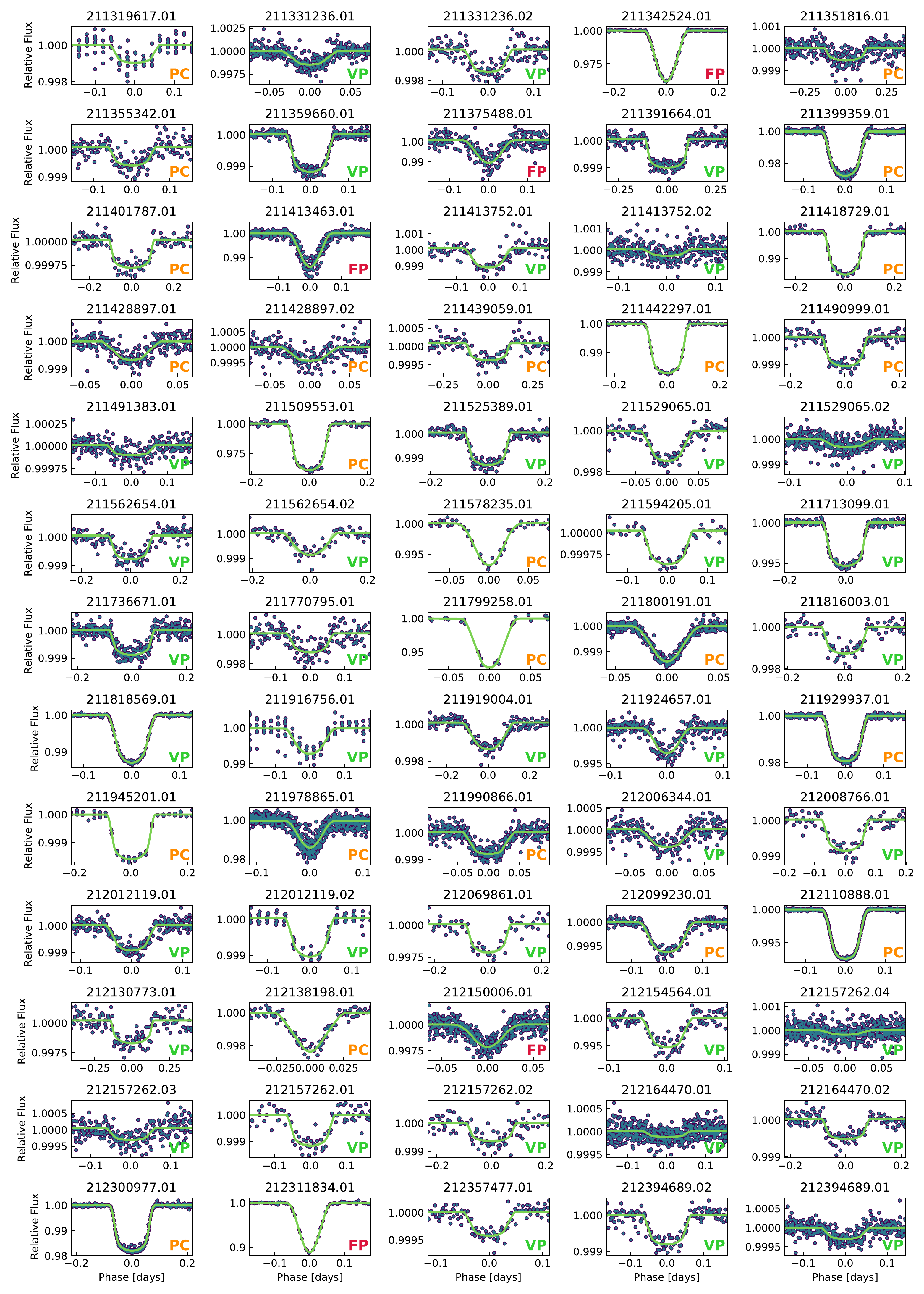}
    \caption{Phase-folded transits (dark blue/green) of validated planets, with the maximum {\it a posteriori} transit models and 1$\sigma$ credible regions (light green) overplotted, and final dispositions in the lower right corner (``VP'' = validated planet (green); ``PC'' = planet candidate (orange); ``FP'' = false positive (red)).}
    \label{fig:plots-1}
\end{figure*}

\begin{figure*}
    \centering
    \includegraphics[width=0.95\textwidth,trim={0.5cm 0.5cm 0.5cm 0.5cm}]{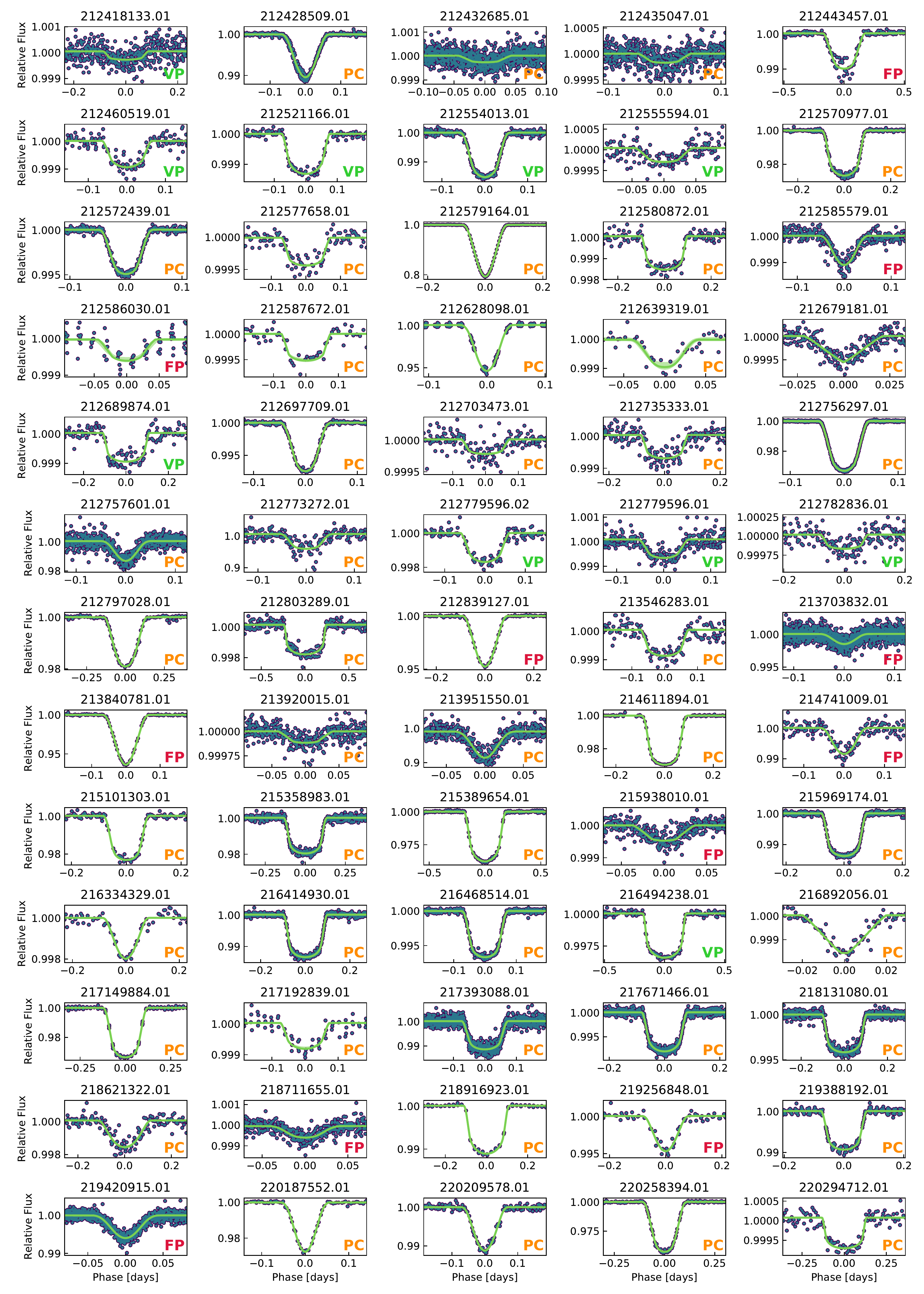}
    \caption{Continuation of \autoref{fig:plots-1}.}
    \label{fig:plots-2}
\end{figure*}

\begin{figure*}
    \centering
    \includegraphics[width=0.95\textwidth,trim={0.5cm 0.5cm 0.5cm 0.5cm}]{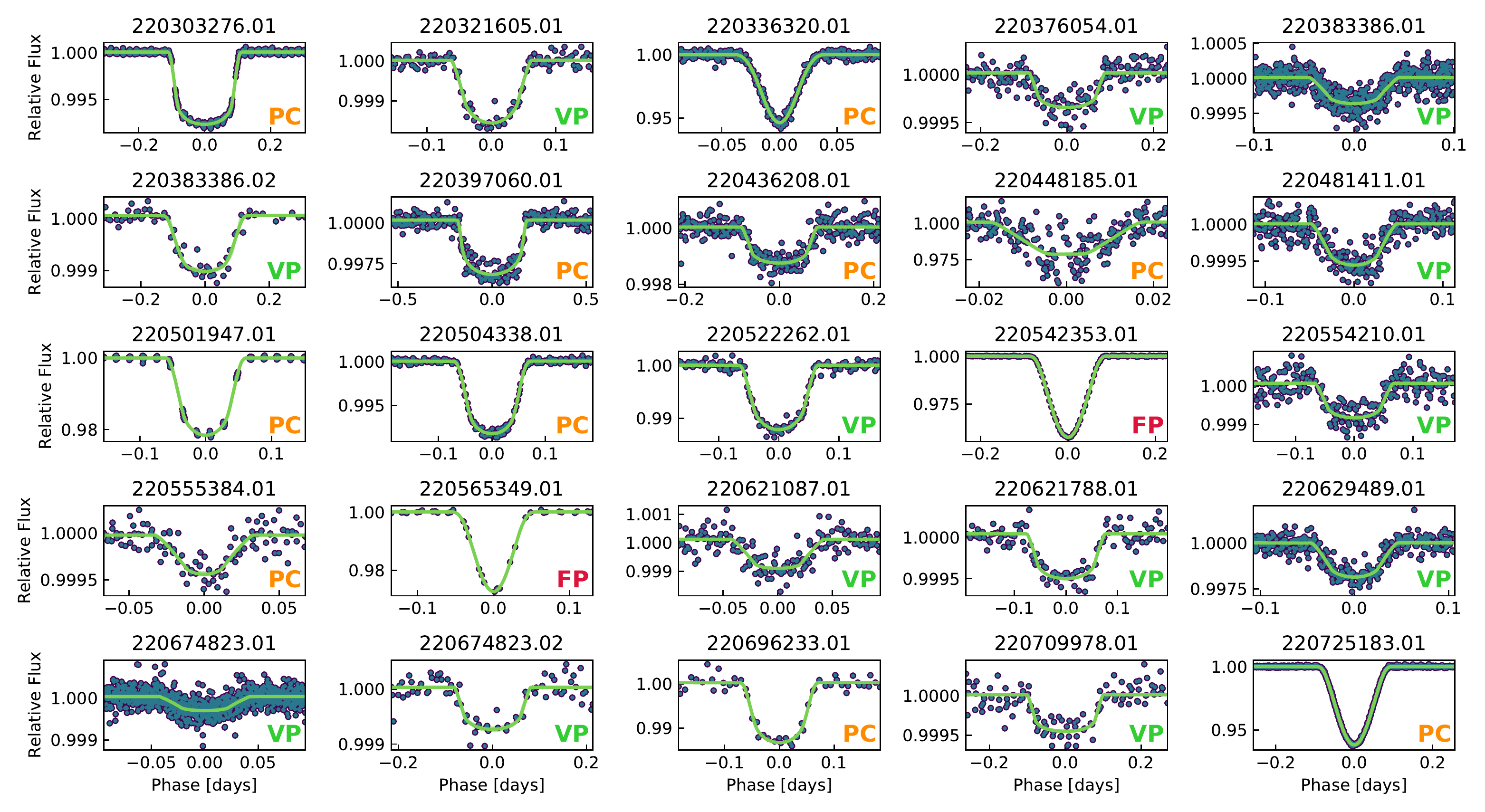}
    \caption{Continuation of \autoref{fig:plots-2}.}
    \label{fig:plots-3}
\end{figure*}

We describe our analysis of the \ktwo light curves in detail in \citet{2016ApJS..226....7C} and P18. In brief, we use MCMC to explore parameter space using the Python packages \texttt{emcee} and \texttt{batman}, an implementation of the analytic light curve model of \citet{2002ApJ...580L.171M}. Parameter estimates resulting from our transit analyses are listed in \autoref{tab:params}. Throughout this work we make heavy use of the Python scientific computing stack (i.e. {\tt numpy}, {\tt scipy}, and {\tt matplotlib}). We plot the phase-folded data and best-fit transit models of each candidate in \autoref{fig:plots-1}, \autoref{fig:plots-2}, and \autoref{fig:plots-3}.

\subsection{Multi-aperture photometry}
\label{sec:multi-aperture}

\autoref{fig:aper1} and \autoref{fig:aper2} show the optimal photometric apertures selected by \ktwophot, which are determined according to an algorithm described in P18. In addition to performing a transit analysis of each candidate using the light curves extracted from these optimal apertures, we also analyzed the light curves produced by circular apertures of varying sizes. This helps to ensure that the transit signal is indeed coming from the target star and not from another nearby source within the aperture, such as the cases identified by \citet{2017A&A...606A..75C}. In such a case, one would expect the measured transit depth to increase as a function of aperture radius. Another possibility is that there is significant photometric dilution from other sources within the aperture, which would result in a decrease in transit depth with larger aperture radius.

To perform this analysis, we extracted light curves using circular apertures with radii of 1.5, 3.0, and 8.0 \kepler pixels (6.0\arcsec, 11.9\arcsec, and 31.8\arcsec, respectively) for each candidate host star. We then fit the transit model to each light curve using the best-fitting parameters from the optimal apertures, with the radius ratio (\Rp/\rstar) allowed to float. To determine the value of \Rp/\rstar and its uncertainty for each aperture extraction, we used the Python package {\tt lmfit}, which utilizes the Levenberg-Marquardt nonlinear least squares minimization in {\tt scipy} \citep{scipy}. By comparing the results from fitting the 1.5, 3.0, and 8.0 pixel radius light curves, we found no evidence of a significant radius dependence for any of the planets we validate in this work (at the 5$\sigma$ level).

Compared to the optimal {\tt k2phot} apertures, these circular apertures typically include substantially different sets of pixels, often resulting in a significantly degraded light curve quality. For some targets this reduces the strength of the constraints from this analysis; a less automated approach and/or a different photometric pipeline could potentially produce better quality light curves for comparison, but this is beyond the scope of this work. We do not validate any systems which exhibit a suspicious radius dependence, or for which nearby bright stars exist but the result of this analysis is ambiguous. For several candidates we classify as false positives, we found evidence of increasing \Rp/\rstar with increasing aperture radius, suggesting that these are actually the eclipses of a nearby EB. For some unvalidated candidates, we found a similar radius dependence, which suggests the signals do not originate from the presumed target star. In some cases we partially rely on the clear absence of a radius dependence to validate a system with a nearby bright star, as the smallest aperture excludes the flux of the neighbor without resulting in a diminished transit depth. We discuss these specific cases at length in \autoref{sec:individual}.

\section{Validation framework}
\label{sec:validation}

This paper represents the final step in a process involving multiple parallel and sequential analyses, and incorporates the results of high resolution imaging and spectroscopic follow-up observations. We rigorously vet the TCEs from \terra to avoid observing stars associated with spurious instrumental signals or obvious astrophysical false positives. The analyses of these follow-up observations then feed into the statistical validation framework described below (see \autoref{fig:flowchart} for an overview). Finally, we take additional steps to ensure the FPPs we compute for each candidate are robust, which we describe in the following subsections.

\subsection{Calculating FPPs}

\begin{figure*}
    \centering
    \includegraphics[width=0.99\textwidth,trim={0.5cm 0 0.5cm 0}]{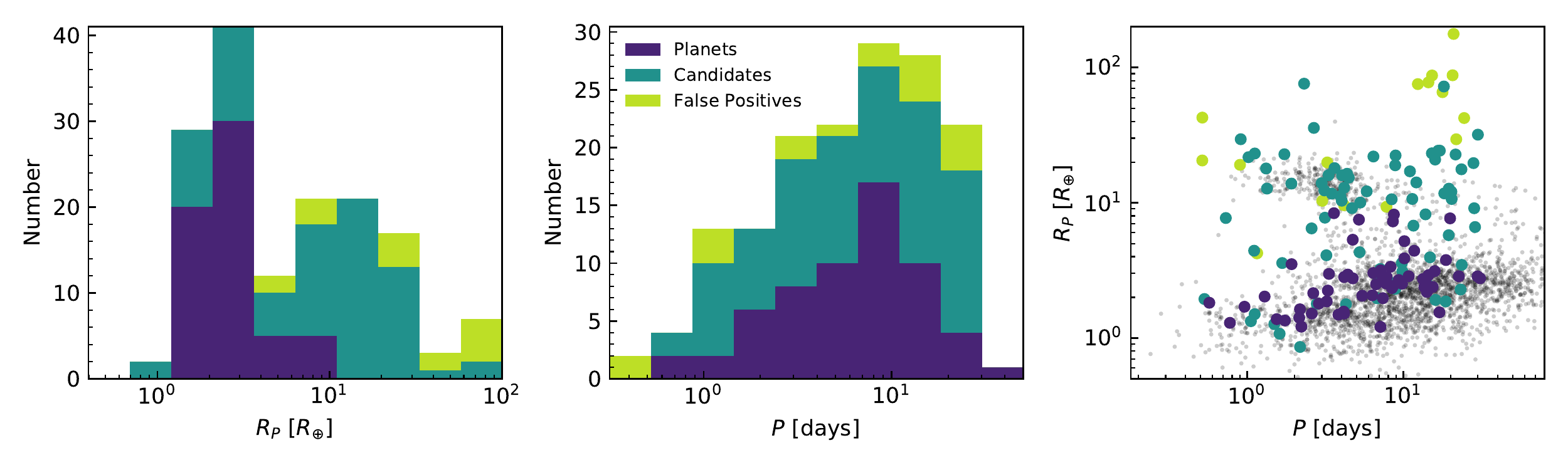}
    \caption{The left and middle panels show stacked histograms of the radius and orbital period distributions of the full sample we present in this work, respectively. The right panel shows radius vs. orbital period, with gray points showing the distribution of previously confirmed planets, based on a query of the NASA Exoplanet Archive on \nexsciquerydate.}
    \label{fig:hist-scatter}
\end{figure*}

To compute FPPs, we use the open-source Python package \vespa. We build off of the methodology of \citet{2016ApJS..226....7C}, who used this approach to compute FPPs for 197 planet candidates from \ktwo's first year (Campaigns C0--C4). The result is a complementary catalog of validated planets, candidates, and false positives for the second year of the \ktwo mission (C5--C8). We adopt the commonly used FPP criteria of \minfpp and \maxfpp for planet validation and false positive designation, respectively (see e.g. \citet{2014ApJ...784...45R,2015ApJ...809...25M,2016ApJ...822...86M}). A candidate with \minfpp $<$ FPP $<$ \maxfpp is designated as neither validated planet nor false positive, and thus remains a planet candidate. \autoref{fig:hist-scatter} shows the distributions of radius and orbital periods for validated planets, candidates, and false positives.

At the heart of \vespa is a robust statistical framework to compute the likelihood of several astrophysical false positive scenarios, both with and without the effect of transit depth dilution caused by additional sources within the photometric aperture -- EBs, hierarchical triple systems (HEBs), and background eclipsing binaries (BEBs). \vespa uses simulations of the galaxy from the \texttt{TRILEGAL} population synthesis code \citep{2005A&A...436..895G}. We emphasize that the inclusion of {\it Gaia} parallaxes significantly impacts the stellar parameters for some candidate host stars, and thus also affects the FPPs; our \vespa results should therefore be more reliable than any previous analyses which did not include parallax. Because \vespa assumes that the input photometry, parallax, and spectroscopic information corresponds to the true host of the transit signals, the FPPs it computes are only reliable when this assumption is valid. We list the likelihoods of false positive scenarios considered by \vespa in \autoref{tab:vespa-cand}.

An important distinction between this work and the validation framework of \citet{2016ApJS..226....7C} is that we have taken extra steps to ensure that our sample of validated planets is pure. Recent work by \citet{2017A&A...606A..75C} and \citet{2017ApJ...847L..18S} showed that several statistically validated planets from \citet{2016ApJS..226....7C} were in fact false positive eclipsing binary scenarios. Therefore, to ensure the high purity of the validated planet sample in this work, we analyzed the light curves from multiple \ktwo apertures, as described in \autoref{sec:modeling}, and also included a planet radius upper limit in our validation criteria, as described in \autoref{sec:large-radii}. These steps ensure that the FPPs are robust for the validated planet sample. However, the unvalidated planet sample contains candidates with low FPPs that we do not validate because of uncertainty about which star is the host; the FPPs for the unvalidated candidate sample are thus inherently less reliable than those of the validated sample, as the assumptions made by \vespa may be sometimes be in violation. We thus urge caution in the interpretation of the FPPs of such unvalidated candidates, in particular for the unvalidated candidates from Campaign 7, whose lower galactic latitudes resulted in frequent contamination from background stars within the photometric apertures. Further observations may help establish which stars are the signal hosts, and could thus enable some of these candidates to be validated or confirmed.

\subsection{Multi-planet systems}

\begin{figure*}
\includegraphics[clip,trim={0.75cm 0cm 1.75cm 0cm},width=0.99\textwidth]{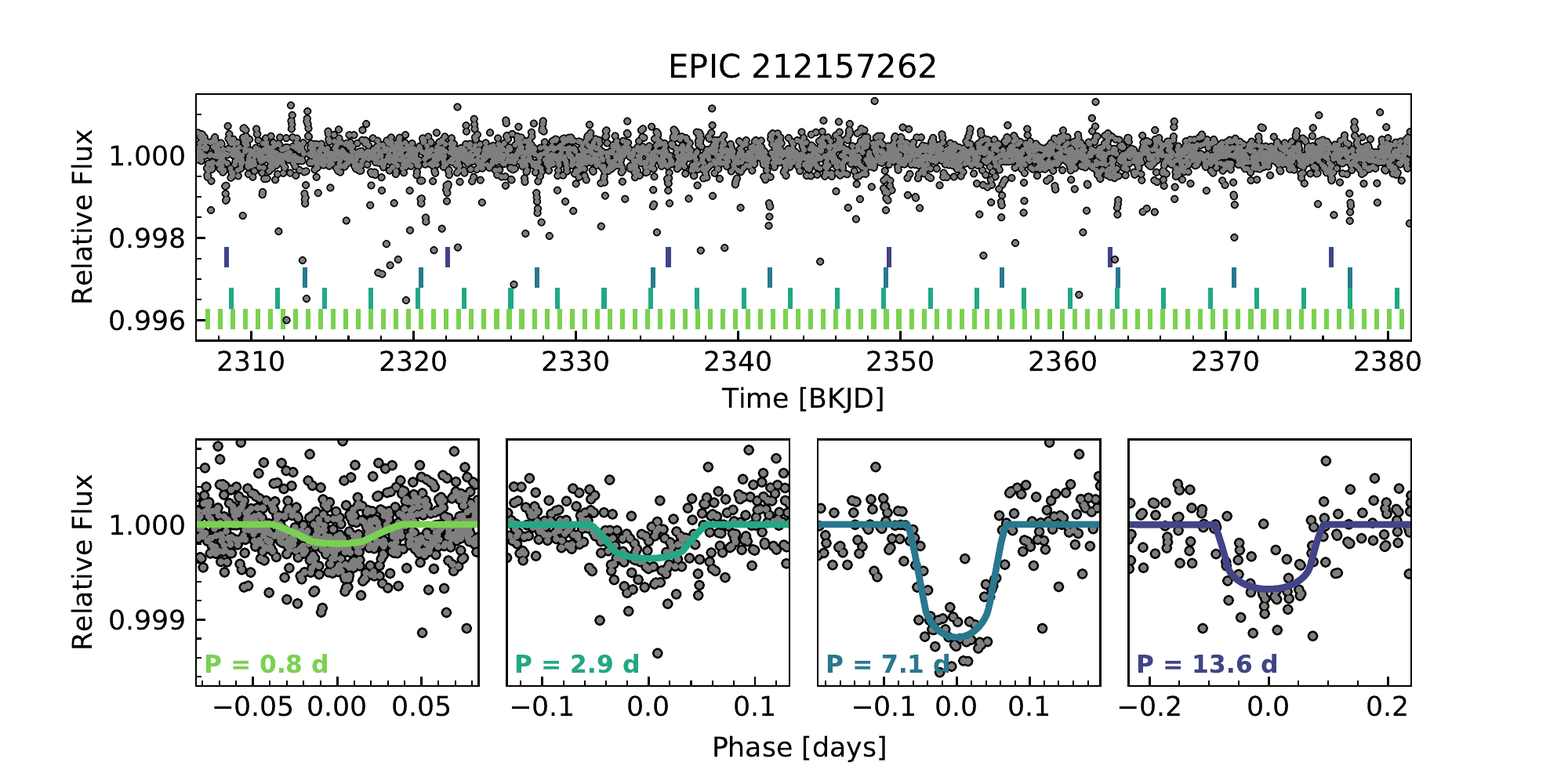}
\caption{K2-187: a validated system of four small planets from Campaign 5. The top panel shows the full \ktwo lightcurve with individual transits indicated by colored tick marks, and the bottom panels show the phase-folded transits of each of the planets, color-coded to match the color of the tick marks.}
\label{fig:212157262}
\end{figure*}

Stars with multiple transiting planet candidates have been shown to have much lower FPPs than their single planet counterparts \citep[e.g.][]{2011ApJS..197....8L,2014ApJ...784...44L}. In this work we take into account the (transiting) planet candidate multiplicity of each system when computing FPPs. Because the FPPs from \vespa do not reflect multiplicity, we apply a ``multiplicity boost'' to the planet scenario, following previous validation papers (e.g. \citealt{2016ApJ...827...78S} and  \citealt{2016ApJS..226....7C}). \citet{2012ApJ...750..112L} estimated this boost factor as 25 for systems of two planet candidates, and a factor of 100 for systems of three or more candidates based on the observed false positive rate for the \kepler field. We apply the boost factors to the planet probability of each member of a multi-candidate system, which reduces the FPP of each individual candidate. \citet{2016ApJ...827...78S} estimated the multiplicity boost for \ktwo using data from fields 1--2 and found values comparable to those found for the \kepler field by \citet{2012ApJ...750..112L}. To check that the factors are not too high for fields 5--8, we used Equations 2 and 4 of \citet{2012ApJ...750..112L} to estimate them from our candidate sample. Based on the FPPs calculated by \vespa, we have a sample purity of $\sim$75\%. In conjunction with the observed fraction of candidates detected in multi-systems (\nmultip/\ntotal), this yields multiplicity boost factors higher than that of the original \kepler field. While the true value of the multiplicity boost is field-dependent, the average values for fields 5--8 appear to be comparable to those of the \kepler field, similar to what was found for fields 1 and 2 by \citet{2016ApJ...827...78S}. Thus, we apply the boost factors estimated by \citet{2012ApJ...750..112L} to all candidates from \ktwo fields 5--8. We note, however, that none of the planets we validate in this work require this boost in order to meet our validation criterion of \minfpp. \autoref{fig:212157262} shows the light curve and phase-folded transits of K2-187, a validated system of four planets detected in Campaign 5.

\subsection{Targets with nearby stellar companions}
\label{sec:companions}

The FPPs we compute with \vespa  are only valid for systems without detected stellar companions within the \ktwo photometric apertures. The high resolution imaging presented in G18 enabled us to detect companions as close as 0.1\arcsec and up to nine magnitudes fainter. If the apparent transit signal originates from a secondary star within the photometric aperture, the large uncertainties on the stellar parameters of the host result in a highly uncertain planetary radius. Furthermore, in most such cases the primary star is much brighter than the host, so the true transit depth would be underestimated by orders of magnitude, making the deep eclipse of a stellar mass object appear more similar to the shallow transit of a planet. We do not validate any planet candidates for which we cannot rule out all detected companions (either from AO or archival imaging) as the source of the signal. To rule out such scenarios, we consider the relationship between the observed transit depth $\delta'$ and the true transit depth $\delta$ given dilution $\gamma$ from a secondary star $\Delta m$ magnitudes fainter than the primary star (in the \kepler bandpass):
\begin{equation}
  \delta' = \frac{\delta}{\gamma} = \frac{\delta}{1 + 10^{0.4\Delta m}}
\end{equation}
To be conservative, we assume a maximum eclipse depth of 100\% (i.e. $\delta = 1$) so if $\delta' > \gamma^{-1}$, then the observed depth is too deep to have originated from the secondary star. Otherwise, the true host of the transit-like signals is uncertain, which in turn induces large uncertainties on the planet radius. \autoref{tab:ao-companions} lists these cases, along with the transit depths and the dilution factors $\gamma_\mathrm{pri}$ and $\gamma_\mathrm{sec}$, which assume the signal originates from the primary or secondary star, respectively. We also indicate these cases of non-validated candidates (regardless of their FPP) by ``AO'' in the Note column of \autoref{tab:params}.

This analysis relies on the results of our extensive high resolution imaging observations (see \autoref{sec:imaging}). The multi-aperture light curve analysis presented in \autoref{sec:modeling} is sensitive to problems of the variety pointed out by \citet{2017A&A...606A..75C}, in which the AO imaging field of view is too small to detect more widely separated stars that nonetheless contribute flux to the \ktwo photometric aperture. However, as described in \autoref{sec:multi-aperture}, the quality of the light curves produced with non-optimal apertures was not always high enough to enable a robust constraint from this analysis. We thus made use of the high precision and completeness of {\it Gaia} DR2 to perform an additional check on the possibility of photometric dilution or false positive contamination from nearby sources.
For each target star, we searched for {\it Gaia} DR2 sources within 2\arcmin\, of the target star positions taken from the EPIC. We then determined the subset of these sources contributing flux to the \ktwo photometric aperture, using a 2-D Gaussian profile to model the point spread function (PSF). According to \kepler documentation\footnote{\url{https://keplergo.arc.nasa.gov/DataAnalysisProducts.shtml}} the full width at half-maximum (FWHM) of the PSF varies from 3.1\arcsec\ to 7.5\arcsec, so we used a value of 6\arcsec as a reasonable approximation for the FWHM across the focal plane. This approach accounts for ``edge cases'' involving a star outside of the aperture, but which still contribute significant flux. Finally, we determined the set of stars contributing enough flux to the aperture to be the source of the observed signals, taking into account dilution from other stars and the observed transit depth (assuming a maximum eclipse depth of 100\%, as before). We show the positions of {\it Gaia} DR2 sources in \autoref{fig:aper1} and \autoref{fig:aper2}, and we color code each {\it Gaia} source according to the following: red squares are sources bright enough (and contributing enough flux) to be the host of the signals, and green squares are sources that are either too faint to be the signal hosts or do not contribute enough flux to the aperture. We indicate cases of multiple stars bright enough to be the host by ``{\it Gaia}'' in the Note column of \autoref{tab:params}.

\subsection{Candidates with large radii}
\label{sec:large-radii}

Several cases of low mass eclipsing stellar companions that were initially classified as planets via statistical validation have recently come to light \citep{2017ApJ...847L..18S}; these stars have radii in the range 0.9--1.9 $R_\mathrm{Jup}\xspace$, consistent with planets in the Jovian size regime (see also \citealt{2012A&A...547A.112M}). Thus, to err on the side of caution, we do not validate any planet candidate with radius larger than 10 \rearth (0.89 \rjup). This cautionary radius threshold can also be seen as an empirically sound choice based on the FPPs of the \kepler candidates presented by \citet{2016ApJ...822...86M}, which rise quickly from $\sim$1\% above 10 \rearth in aggregate (see e.g. Figure 4 of \citealt{2016ApJ...822...86M}). Candidates with radii larger than this in \autoref{tab:params} are indicated by ``LR'' in the Note column. Many candidates with large radii are clear false positives based on their FPPs, but we do not validate several candidates with large radii in spite of their low FPPs. Future RV measurements of these unvalidated candidates are likely to reveal many of them to be giant planets.

\section{Discussion}
\label{sec:discussion}

We present the candidates, dispositions, and parameters of interest in \autoref{tab:params}. Based on a query of the NASA Exoplanet Archive\footnote{\url{https://exoplanetarchive.ipac.caltech.edu/} ``Confirmed Planets'' table queried on \nexsciquerydate} \citep{2013PASP..125..989A}, \nalreadyvalid of our validated planets have already been statistically validated or confirmed via RV measurements. This leaves a remainder of \nnewlyvalid newly validated planets, \ncandidates unvalidated candidates, and \nfp false positives. None of the false positives we identified have been presented as validated planets in the literature, and neither are we aware of a false positive designation in the literature for any of the planets in our validated sample. The distributions of radius and orbital period are shown in \autoref{fig:hist-scatter}, colored by their final disposition. We see a tendency for false positives to have large radii, which indicates that most of them are the result of eclipsing binary scenarios with little to no dilution from blended stars. We present notes on several individual systems in \autoref{sec:individual}.

In \autoref{fig:enhancement}, we show the impact of this work by plotting the fractional enhancement to the previously known planet population. as a function of radius and host star magnitude. The enhancement from \ktwo C5--8 is similar to the enhancement from \ktwo C0--4 (see \citet{2016ApJS..226....7C}); planets with smaller radii and brighter host stars are particularly enhanced by this study. This is the result of the larger number of nearby and/or later spectral type stars observed by \ktwo as compared to \kepler, which is primarily due to the community-driven target list and wider survey of \ktwo. Indeed, the median optical and $J$-band magnitudes of \ktwo planet hosts are $\sim$1.9 and $\sim$2.1 mag brighter than \kepler planet hosts, respectively. We expect this trend to continue for the remainder of \ktwo and soon also with \tess \citep{2014SPIE.9143E..20R}.

While this paper was in preparation, \citet{2018AJ....155..136M} published an independently produced catalog for Campaigns 0--10. We note that our catalog contains an additional number of validated planets roughly equal to the number of validated planets in common, due to a combination of differences in the light curve extractions, follow-up observations and analyses, validation criteria, and the limitation of \citet{2018AJ....155..136M} to relatively bright host stars ($Kp < 13$ mag).
Besides using different light curve extractions and including some fainter host stars, our validated sample has the following differences: our validation threshold is less conservative (1\% vs 0.1\%); we incorporate FPP constraints derived from high resolution spectroscopy and {\tt ReaMatch} (see \autoref{sec:stellar}); we utilize a larger set of contrast curve constraints from AO/speckle imaging (\citet{2018AJ....155..136M} utilized only the contrast curves publicly available on ExoFOP at the time of submission). This underscores the utility of multiple teams conducting follow-up observations and independent analyses.

\subsection{Interesting systems}

Our validated planet sample includes \nsmall planets smaller than 2 \rearth, several of which orbit bright host stars ($J$$\approx$7--9). These small planets thus present opportunities for detailed studies of terrestrial worlds, either via RV measurements or transmission spectroscopy. Our sample also contains \nmultip validated planets in \nmultis multi-systems, some of which orbit near low-order mean motion resonances. In particular, further transit monitoring could reveal TTVs in systems such as EPIC\,211562654\,bc, which is within 5\% of a 2:1 period commensurability. This growing population of resonant systems (e.g. TRAPPIST-1 \citep{2017Natur.542..456G} and K2-138 \citep{2018AJ....155...57C}) provides important clues for planet formation theories.

Also present in the validated planet sample are \nusp planets with periods less than one day, commonly referred to in the literature as ultra-short period planets \citep[USPs; e.g.][]{2013ApJ...774...54S}. These planets may have migrated to their current orbital locations, or their orbits could be the result of a scattering event followed by tidal circularization. In particular, we validate the previously confirmed USP K2-96 b \citep[HD 3167 b;][]{2016ApJ...829L...9V, 2017AJ....154..122C}, which is part of a system with a rich dynamical history. We also validate the previously confirmed USP K2-106\,b \citep[EPIC\,220674823\,b;][]{2017AJ....153...82A, 2017AJ....153..271S}, which is part of a multi-planet system. The validated four-planet system in our sample K2-187 also contains a USP, making it a potentially interesting system from a dynamical point of view. However, given the low mutual inclinations implied by the presence of four transiting planets in this system, it is likely that the system has a quiet dynamical history, perhaps similar to what has been seen in other multi-planet USP-hosting systems, such as WASP-47 \citep{2017AJ....153...70S, 2017AJ....154..237V}.

\subsection{RV targets}

To identify compelling targets for future characterization studies, we predicted masses using the mass-radius relation of \citep{2016ApJ...825...19W} and then used these to predict RV semi-amplitudes. The following validated planets are the top three most compelling targets for future RV mass measurements, in the sense that they orbit relatively bright stars ($J < 10$, $V < 10.7$), have expected semi-amplitudes within reach of current and planned precision spectrographs ($K_\mathrm{pred} > 1$ m/s), and currently lack a mass determination: 211594205.01 (K2-184\,b), 212357477.01 (K2-277\,b), and 220709978.01 (K2-222\,b).
Additionally, these are interesting targets because their radii place them near the recently observed gap in the radius distribution \citep{2017AJ....154..109F, 2018MNRAS.479.4786V, 2018arXiv180501453F}. Precise mass measurements would yield planet densities and therefore provide insights into the possibility of having been sculpted by photoevaporation \citep[e.g.][]{2013ApJ...775..105O, 2014ApJ...792....1L}. Such tests of photoevaporation theories will in turn help to clarify the extent to which the diversity of planet densities can be explained by planetesimal accretion \citep{2016ApJ...817L..13I} or core-powered mass-loss \citep{2018MNRAS.476..759G}, as well as probing core compositions \citep{2018ApJ...853..163J}.

\subsection{Previously validated planets}

Of the \nvalidated validated planets in our sample, \nalreadyvalid have previously been validated in the literature, and we include the default names of these planets as they are listed in the NASA Exoplanet Archive alongside their candidate IDs in \autoref{tab:params}.
We found overall good agreement with literature parameter estimates, but some discrepancies likely attributable to some combination of differences in stellar characterization and photometric extraction, time-series detrending, and transit-fitting. One interesting example is K2-97\,b, for which our estimate of \Rp/\rstar is $\sim$3$\sigma$ different from the value found by \citet{2016AJ....152..185G}; K2-97 is an evolved star exhibiting asteroseismic oscillations detectable in the \ktwo photometry.

\begin{figure*}
\includegraphics[clip,trim={0.5cm 0cm 0.25cm 0cm},width=0.99\textwidth]{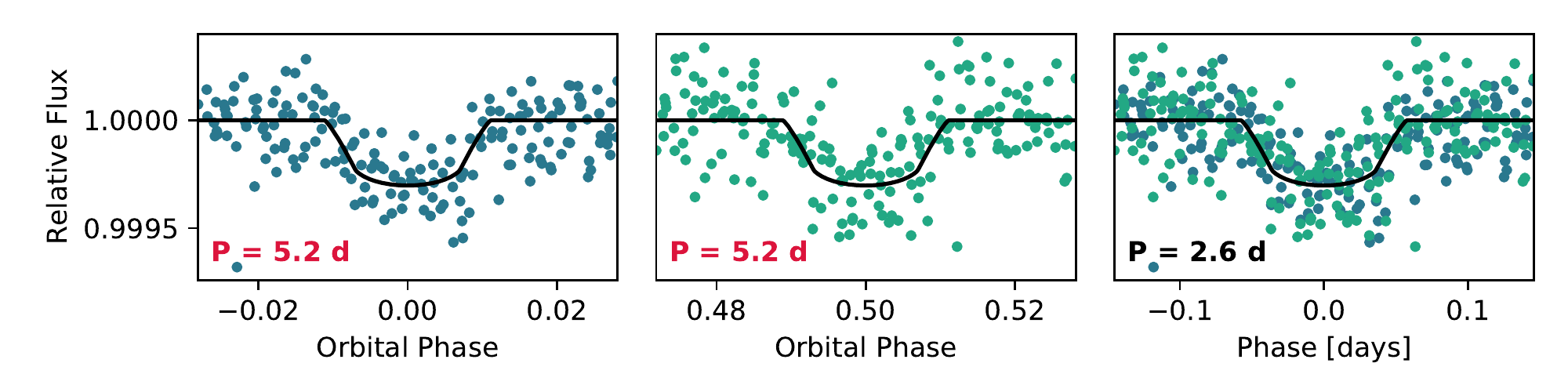}
\caption{Left and middle panels: \ktwo light curve of K2-189\,b, folded on the period reported by \citet{2018AJ....155..136M}. The left panel shows the folded light curve centered at orbital phase 0.0 (blue), and the middle plot shows the same folded light curve centered at phase 0.5 (green). Right panel: \ktwo light curve of K2-189\,b folded on the best fit period from our analysis, with the data points color coded to match their appearance in the left and middle panels. Overplotted on the data in each panel is the best fit transit model from our analysis (black).}
\label{fig:4689}
\end{figure*}

\begin{figure}
    \centering
    \includegraphics[width=0.49\textwidth,trim={0.5cm 0 0 0}]{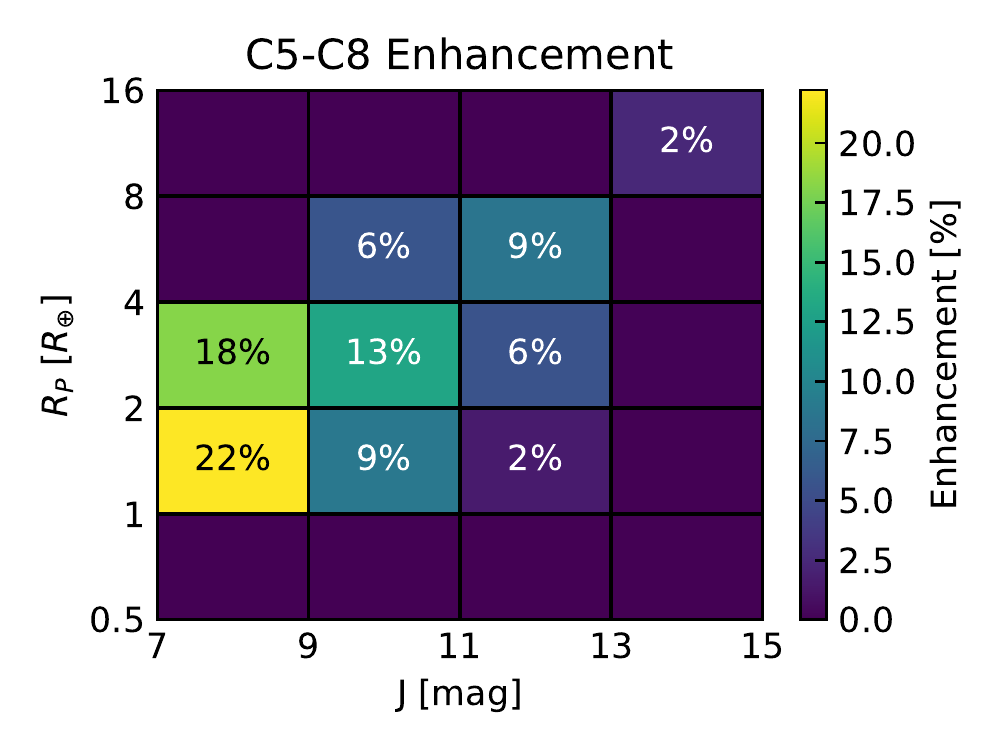}
    \caption{Fractional enhancement (in percent) to the population of known planets from \ktwo C5-C8, based on the newly validated planets from this work as compared to the previously confirmed planets in the NASA Exoplanet Archive on \nexsciquerydate.}
    \label{fig:enhancement}
\end{figure}

Besides these discrepancies with the parameter estimates from the literature, we found one case of system parameters arising from the mis-identification of a candidate's orbital period. \citet{2018AJ....155..136M} recently reported two small validated planets orbiting K2-189 (EPIC\,212394689), which is also a star in our sample. We also validated two small planets orbiting this star, but we found that the orbital period of the inner planet reported by \citet{2018AJ....155..136M} is twice the true period. To check the validity of this conclusion, we fitted the light curve at both the period we detected and the period reported by \citet{2018AJ....155..136M}. We then visually inspected the fits, as well as compared the resulting fit parameters. Our estimates of the planet-to-star radius ratio (\Rp/\rstar) are in 1$\sigma$ agreement, but our estimate is the larger of the two. If the true orbital period were in fact twice our estimate, one would expect our estimate of \Rp/\rstar to be significantly smaller due to the presence of out-of-transit photometry at the location of every other presumed transit. When we fold the light curve on the period reported by \citet{2018AJ....155..136M}, we find that our best-fit model is a good fit to the data at both phase 0.0 and at phase 0.5, indicating that the data were folded on twice the true period (see \autoref{fig:4689}). We also analyzed the light curve analyzed by \citet{2018AJ....155..136M}, which was produced by K2SFF \citep{2014PASP..126..948V} and is publicly available\footnote{\url{https://archive.stsci.edu/prepds/k2sff/}}; however, we came to the same conclusion.
Besides the case of K2-189\,b, there is good agreement in planet parameters ($< 3 \sigma$) for the other planets in our validated sample which are also validated by \citet{2018AJ....155..136M}.

\subsection{\ktwo self-follow-up}

\ktwo C16 observed a field overlapping C5 from December 7, 2017, to February 25, 2018. We used {\tt kadenza}\footnote{\url{https://github.com/KeplerGO/kadenza}} to process the \ktwo raw cadence data and then analyzed the resulting target pixel files with our team's standard pipelines, as described in detail in \citet{2018arXiv180304091Y}. A subset of our C5 candidates were also observed by \ktwo during C16, so to demonstrate the increase in precision of orbital period estimates from a second observing campaign 18 months later, we conducted a joint analysis of both the C5 and C16 light curves for this subset of targets.\footnote{In addition to our C5--8 light curves, our C16 light curves are also publicly available on ExoFOP: \url{https://exofop.ipac.caltech.edu}} \autoref{tab:c16} lists the resulting period estimates, along with the C5-only estimates for comparison. We find a median improvement of $\sim$26X. This illustrates the utility of follow-up transit observations; such improvements in ephemeris estimates greatly facilitate efficient scheduling of atmospheric transmission spectroscopy with expensive telescope assets, e.g. \jwst. For deep enough transits, similar improvements are possible with ground-based transit follow-up observations, but for many interesting targets such follow-up will need to be conducted from space, e.g. with \spitzer (see Livingston et al., in review) or \cheops. A similar overlap exists between \ktwo C6 and C17, and this type of space-based self-follow-up will likely to continue with the \tess extended mission.

\subsection{\ktwo yield}

The newly validated planets in this work bring the total number of planets discovered by \ktwo to over $\sim$360.\footnote{See footnote 5} Our sample of \ntotal candidates has an integrated FPP of $\sim$35, and we identified \nfp of these as false positives; thus 17 additional false positives can reasonably expected to be found among the \ncandidates unvalidated candidates. By extrapolation of the true positive rate implied by our integrated FPPs for C5--C8, we expect that \ktwo will have discovered $\sim$600 planets by the end of 2018 (i.e. by the end of Campaign 19), which is approximately when \ktwo will run out of the fuel required for 3-axis pointing control \citep{2014PASP..126..398H}. Many of these planets will remain as unvalidated candidates until sufficient follow-up observations can be made. Many are potential targets for RV mass measurement using current and upcoming high precision spectrographs in the optical and near IR, especially given the relatively bright host stars typically surveyed by \ktwo. These planets are complementary to those expected to be discovered by the upcoming NASA \tess mission, due to the fact that \ktwo surveys the ecliptic plane and \tess will survey most of the remainder of the celestial sphere.

\section{Summary}
\label{sec:summary}

We have synthesized \ktwo light curve analyses, high resolution spectroscopic and imaging host star characterization, and false positive probabilities for \ntotal planet candidates identified in fields C5--C8, resulting in \nvalidated statistically validated planets. We identified \nfp false positives from among our candidate sample, leaving a remainder of \ncandidates unvalidated candidates, most of which are likely to be real planets that future observations and analyses could potentially validate. Of the \nvalidated validated planets, \nnewlyvalid are new discoveries, and some are potential targets for future study with high precision radial velocity instruments and \jwst. \ktwo's transit survey of the ecliptic plane is complementary to the upcoming NASA \tess mission, and the utilization of telescope resources among large collaborations such as required for this work foreshadows the necessity for coordinated and efficient teamwork in the \tess era.

\acknowledgements
This paper includes data collected by the K2 mission. Funding for the K2 mission is provided by the National Aeronautics and Space Administration (NASA) Science Mission directorate. This work benefited from the Exoplanet Summer Program in the Other Worlds Laboratory (OWL) at the University of California, Santa Cruz, a program funded by the Heising-Simons Foundation. Portions of this work were performed under contract with the Jet Propulsion Laboratory (JPL) funded by NASA through the Sagan Fellowship Program executed by the NASA Exoplanet Science Institute. This research has made use of the NASA Exoplanet Archive, which is operated by Caltech, under contract with NASA's Exoplanet Exploration Program. This work made use of the SIMBAD database (operated at CDS, Strasbourg, France) and NASA’s Astrophysics Data System Bibliographic Services. This research made use of the Infrared Science Archive, which is operated by the California Institute of Technology (Caltech), under contract with NASA. Portions of this work were performed at Caltech under contract with NASA. J.\,H.\,L. gratefully acknowledges the support of the Japan Society for the Promotion of Science (JSPS) Research Fellowship for Young Scientists. E.\,S. is supported by a postgraduate scholarship from the Natural Sciences and Engineering Research Council of Canada. E.\,A.\,P. acknowledges support by NASA through a Hubble Fellowship grant awarded by the Space Telescope Science Institute, which is operated by the Association of Universities for Research in Astronomy, Inc., for NASA, under contract NAS 5-26555. B.\,J.\,F. was supported by the National Science Foundation Graduate Research Fellowship under grant No. 2014184874. A.\,W.\,H. acknowledges support for our K2 team through a NASA Astrophysics Data Analysis Program grant. A.\,W.\,H. and I.\,J.\,M.\,C. acknowledge support from the K2 Guest Observer Program. This work has made use of data from the European Space Agency (ESA) mission {\it Gaia} (\url{https://www.cosmos.esa.int/gaia}), processed by the {\it Gaia} Data Processing and Analysis Consortium (DPAC, \url{https://www.cosmos.esa.int/web/gaia/dpac/consortium}). Funding for the DPAC has been provided by national institutions, in particular the institutions participating in the {\it Gaia} Multilateral Agreement. Some of the data presented herein were obtained at the W.M. Keck Observatory, which is operated as a scientific partnership between Caltech, the University of California, and NASA. The authors wish to extend special thanks to those of Hawai'ian ancestry, on whose sacred mountain of Maunakea we are privileged to be quests. We are most fortunate to have the opportunity to conduct observations from this mountain.

\facilities{\kepler, {\it Gaia}, Keck (NIRC2, HIRES), Gemini (DSSI, NIRI), Palomar (PHARO, TripleSpec), WIYN (NESSI), NTT (SOFI), IRTF (SpeX)}

\software{{\tt numpy} \citep{numpy}, {\tt scipy} \citep{scipy}, {\tt matplotlib} \citep{Hunter:2007}, {\tt lmfit} \citep{newville_2014_11813}, {\tt emcee} \citep{emcee}, {\tt batman} \citep{2015PASP..127.1161K}, {\tt isochrones} \citep{2015ascl.soft03010M}, {\tt vespa}, \citep{2015ascl.soft03011M}}

\bibliography{ref.bib}

\appendix

\startlongtable
\begin{deluxetable*}{lllllllcr}
\tabletypesize{\tiny}
\setlength\tabcolsep{4pt}
\tablecaption{Target stars and parameters, where `Provenance' indicates the basis of the parameters (see \autoref{sec:stellar}): SMs = {\tt SpecMatch-syn}, SMe = {\tt SpecMatch-emp}, D17 = \citet{2017ApJ...836..167D}, M17 = \citet{2017ApJ...837...72M}, and JHK = 2MASS photometry, and the \ktwo GO programs associated with each star are listed in the final column. \label{tab:stellar}}
\tablehead{EPIC & $T_{\mathrm{eff}}$ & \logg & [Fe/H] & \mstar & \rstar & Distance & Provenance & GO programs\\
       & [K] & [cgs] & [dex] & [$M_{\odot}$] & [$R_{\odot}$] & [pc] & & }
\startdata
211319617  &   $5358 \pm 65$  &  $4.58 \pm 0.03$  &  $-0.54 \pm 0.06$  &  $0.72 \pm 0.03$  &     $0.72 \pm 0.01$  &     $204.51 \pm 1.70$  &  SMsyn  & 5007,5029,5060,5106 \\ 
211331236  &   $3778 \pm 53$  &  $4.75 \pm 0.01$  &  $-0.05 \pm 0.13$  &  $0.54 \pm 0.01$  &     $0.51 \pm 0.01$  &     $101.95 \pm 0.37$  &    D17  & 5020 \\ 
211342524  &   $6174 \pm 89$  &  $4.05 \pm 0.03$  &  $-0.15 \pm 0.06$  &  $1.21 \pm 0.04$  &     $1.72 \pm 0.05$  &    $606.14 \pm 15.04$  &  SMsyn  & 5007,5029,5033 \\ 
211351816  &   $4717 \pm 73$  &  $3.22 \pm 0.04$  &   $0.33 \pm 0.04$  &  $1.20 \pm 0.05$  &     $4.47 \pm 0.18$  &    $869.76 \pm 31.37$  &  SMsyn  & 5020,5089 \\ 
211355342  &   $5608 \pm 91$  &  $4.38 \pm 0.03$  &   $0.23 \pm 0.06$  &  $0.99 \pm 0.04$  &     $1.07 \pm 0.02$  &     $364.82 \pm 5.36$  &  SMsyn  & 5007,5029,5033,5060,5106 \\ 
211359660  &   $5165 \pm 50$  &  $4.58 \pm 0.01$  &   $0.09 \pm 0.05$  &  $0.88 \pm 0.02$  &     $0.80 \pm 0.01$  &     $155.90 \pm 1.19$  &  SMsyn  & 5007,5029,5033,5060,5104,5106 \\ 
211375488  &  $3707 \pm 484$  &  $4.77 \pm 0.16$  &   $0.01 \pm 0.16$  &  $0.51 \pm 0.18$  &     $0.49 \pm 0.17$  &   $707.82 \pm 393.96$  &    JHK  & 5096 \\ 
211391664  &   $6172 \pm 82$  &  $4.13 \pm 0.02$  &  $-0.06 \pm 0.05$  &  $1.20 \pm 0.03$  &     $1.57 \pm 0.04$  &    $526.31 \pm 11.13$  &  SMsyn  & 5007,5029 \\
\enddata
\end{deluxetable*}

\begin{longrotatetable}
\begin{deluxetable*}{lcccccccccccccc}
\tabletypesize{\tiny}
\setlength\tabcolsep{1pt}
\tablecaption{Candidate parameters and dispositions (``VP'' = validated planet; ``PC'' = planet candidate; ``FP'' = false positive). \label{tab:params}}
\tablehead{Candidate & Field & Name & $P$ & $T_0$ & \Rp/\rstar & $a$/\rstar & $b$ & \Rp & $a$ & $T_\mathrm{eq}$ & $\rho_\star$ & FPP & Disp. & Note \\
 & & & [days] & [BKJD] & [\%] &  &  & [\rearth] & [AU $\times 10^{-2}$] & [K] & [$\rho_\odot$] &  &  & }
\startdata
211319617.01     &         5     &          K2-180 b     &   $8.867209^{+0.001068}_{-0.001016}$     &  $2310.38934^{+0.00448}_{-0.00453}$     &     $2.94^{+0.24}_{-0.17}$     &    $24.12^{+2.86}_{-6.05}$     &  $0.42^{+0.32}_{-0.28}$     &       $2.31 \pm 0.16$     &   $7.53^{+0.10}_{-0.10}$     &       $731^{+11}_{-11}$     &       $2.394^{+1.450}_{-1.037}$     &   $8.9 \times 10^{-6}$     &          PC     &            Gaia  \\
211331236.01     &         5     &          K2-117 b     &   $1.291600^{+0.000049}_{-0.000051}$     &  $2309.77452^{+0.00162}_{-0.00158}$     &     $3.63^{+0.21}_{-0.14}$     &     $7.87^{+0.81}_{-1.77}$     &  $0.39^{+0.31}_{-0.27}$     &       $2.03 \pm 0.10$     &   $1.89^{+0.01}_{-0.01}$     &       $868^{+14}_{-14}$     &       $3.934^{+2.046}_{-1.521}$     &  $2.5 \times 10^{-10}$     &          VP     &              MS  \\
211331236.02     &         5     &          K2-117 c     &   $5.444017^{+0.000462}_{-0.000455}$     &  $2310.56630^{+0.00358}_{-0.00352}$     &     $3.68^{+0.24}_{-0.19}$     &    $18.52^{+2.17}_{-4.33}$     &  $0.40^{+0.32}_{-0.28}$     &       $2.06 \pm 0.12$     &   $4.92^{+0.03}_{-0.03}$     &         $537^{+9}_{-9}$     &       $2.879^{+1.669}_{-1.200}$     &   $6.7 \times 10^{-8}$     &          VP     &              MS  \\
211342524.01     &         5     &                       &  $14.448770^{+0.000110}_{-0.000111}$     &  $2316.36774^{+0.00027}_{-0.00026}$     &   $41.31^{+10.37}_{-9.91}$     &    $25.00^{+0.28}_{-0.19}$     &  $1.13^{+0.12}_{-0.12}$     &     $77.60 \pm 19.11$     &  $12.36^{+0.14}_{-0.14}$     &      $1016^{+22}_{-21}$     &       $1.004^{+0.028}_{-0.028}$     &                      1     &          FP     &              LR  \\
\enddata
\tablecomments{AO -- bright nearby star detected in AO; {\it Gaia} -- bright star within \ktwo aperture detected in {\it Gaia} DR2; LR -- suspiciously large radius; MS -- multi-system; TTV -- flagged for TTVs; USP -- ultra-short period.}
\end{deluxetable*}
\end{longrotatetable}

\begin{deluxetable}{lcccc}
\tabletypesize{\tiny}
\tablecaption{Nearby bright AO sources. \label{tab:ao-companions}}
\tablehead{Primary & $\rho$ & $\Delta Kp$ & $\gamma_{\mathrm{pri}}$ & $\gamma_{\mathrm{sec}}$ \\
 EPIC ID & [arcsec] & [mag] & & }
\startdata
212012119 & 7.11 & 2.3 & 1.12 & 9.32 \\
211439059 & 0.23 & 0.1 & 1.91 & 2.10 \\
211428897 & 1.08 & 0.8 & 1.48 & 3.09 \\
211490999 & 4.67 & 5.4 & 1.01 & 145.54 \\
211413752 & 4.73 & 5.9 & 1.00 & 230.09 \\
212150006 & 0.35 & 0.9 & 1.44 & 3.29 \\
211509553 & 2.00 & 3.3 & 1.05 & 21.89 \\
212703473 & 0.21 & 0.1 & 1.91 & 2.10 \\
212577658 & 1.42 & 0.4 & 1.69 & 2.45 \\
212735333 & 9.29 & 3.7 & 1.03 & 31.20 \\
212679181 & 1.24 & 0.3 & 1.76 & 2.32 \\
212555594 & 13.55 & 2.8 & 1.08 & 14.18 \\
212572439 & 6.12 & 0.8 & 1.48 & 3.09 \\
212432685 & 0.66 & 0.9 & 1.44 & 3.29 \\
212628098 & 1.88 & 1.1 & 1.36 & 3.75 \\
213920015 & 1.08 & 0.1 & 1.91 & 2.10 \\
218711655 & 4.28 & 4.1 & 1.02 & 44.65 \\
214611894 & 1.13 & 1.8 & 1.19 & 6.25 \\
218621322 & 8.10 & 3.0 & 1.06 & 16.85 \\
217192839 & 6.56 & 6.7 & 1.00 & 479.63 \\
218131080 & 0.21 & 1.4 & 1.28 & 4.63 \\
219420915 & 11.17 & 3.0 & 1.06 & 16.85 \\
219256848 & 2.94 & 1.2 & 1.33 & 4.02 \\
215389654 & 8.13 & 2.9 & 1.07 & 15.45 \\
216414930 & 1.50 & 3.7 & 1.03 & 31.20 \\
215358983 & 3.41 & 4.2 & 1.02 & 48.86 \\
213703832 & 0.37 & 3.9 & 1.03 & 37.31 \\
215101303 & 8.53 & 2.3 & 1.12 & 9.32 \\
220294712 & 1.34 & 5.5 & 1.01 & 159.49 \\
220555384 & 0.20 & 0.2 & 1.83 & 2.20 \\
220397060 & 1.48 & 3.9 & 1.03 & 37.31 \\
220436208 & 6.48 & 3.2 & 1.05 & 20.05 \\

\enddata
\tablenotetext{}{Only AO sources bright enough to produce the observed transit-like signals are listed. $\gamma_{\mathrm{pri}}$ and $\gamma_{\mathrm{sec}}$ are the dilution factors, assuming the transit signal comes from the primary and secondary stars, respectively.}
\end{deluxetable}

\section{Notes on individual systems}
\label{sec:individual}

The multi-candidate systems EPIC\,212012119 and EPIC\,212779596 both have bright nearby stars that could potentially make the origin of the signals unclear (see \autoref{tab:params}). EPIC\,212012119 hosts candidates with radii of $2.24 \pm 0.12$ and $2.34 \pm 0.10$ \rearth, and orbital periods of $3.2810 \pm 0.0001$ and $8.4388 \pm 0.0003$ days, respectively; EPIC\,212012084 is $\sim$7\arcsec\, away and 3.2 magnitudes fainter. Similarly, EPIC\,212779596 hosts candidates with radii of $1.87 \pm 0.14$ and $2.78 \pm 0.13$ \rearth, and orbital periods of $3.2253 \pm 0.0002$ and $7.3747 \pm 0.0004$ days, respectively; EPIC\,212779556 is $\sim$8\arcsec\, away and 5.6 magnitudes fainter. The neighboring stars can be seen by eye in the Pan-STARRS-1 $r$-band images, and are clearly within the overplotted {\tt k2phot} apertures (see \autoref{fig:aper1} and \autoref{fig:aper2}). However, because they are separated by several \kepler pixels from their respective primary stars, we can use multi-aperture photometry to identify the source of the observed signals (see \autoref{sec:modeling}). For both systems, this analysis suggests that the signals originate from the primary stars -- the smaller apertures exclude most of the photons from the nearby stars, but there is no apparent radius dependence of transit depth. Furthermore, it is {\it a priori} likely that each pair of candidates orbits the same star, and if the planets orbit the fainter star, the true radii would be significantly larger because of dilution from the primary. If we assume the candidates associated with EPIC\,212012119 are in fact transiting the fainter star, then the implied planet radii would be $\sim 5$ times larger than Jupiter (assuming EPIC\,212012084 has a similar radius). It is far more plausible that EPIC\,212012119 hosts two sub-Neptunes. Similarly, it is more likely that two sub-Neptunes transit the star EPIC\,212779596 than a scenario in which the signals are caused by eclipses of the fainter star by stellar-sized objects. Thus, we conclude that these are both valid multi-planet systems. Because of dilution from the secondary star, the radii of EPIC\,212012119\,bc are potentially larger than we report by up to 4\%, but this is within the error bars. In the case of EPIC\,212779596\,bc the dilution from the secondary star is negligible, at less than 1\%.

The star EPIC\,211428897 has spectral type M2V \citep{2017ApJ...836..167D} and hosts two apparent sub-Earth-sized planets on 1.6 and 2.2 day orbits, but a nearby star detected in AO and speckle imaging complicates the interpretation of this system. The companion is separated by 1.1\arcsec\, and is fainter by $\sim$1.8 and $\sim$1.2 magnitudes in the (approximate) $r$ and $z$ band filters used by DSSI and NESSI. A priori, the close separation and color of this star suggest that it is a bound late-type companion, and the galactic latitude (b = 28.48 degrees) implies only modest levels of contamination by background giants. Intriguingly, two sources are listed in {\it Gaia} DR2 near the position of EPIC\,211428897, with separations of 0.86\arcsec and 1.92\arcsec, respectively. The first star ({\it Gaia} DR2 ID 602557012250320768) is listed with a $G$-band magnitude of 14.52 but no parallax or proper motion. The second star ({\it Gaia} DR2 ID 602557012249101696) is listed with a $G$-band magnitude of 13.31, a parallax of 20.99$\pm$0.10 mas, and a proper motion of $\mu_\alpha$ = -62.58$\pm$0.16 mas, $\mu_\delta$ = -104.03$\pm$0.10 mas. Using the values of \teff, \mstar, \rstar, and \feh reported by \citet{2017ApJ...836..167D} as spectroscopic priors, we use the {\it Gaia} DR2 parallax and 2MASS $JHK$ photometry to estimate a distance of 47.5$\pm$0.2 pc via the {\tt isochrones} package. As expected, there is no indication of a dependence of transit depth on aperture size, due to the proximity of the companion, which is well within the smallest aperture. Accounting for dilution and assuming similar stellar radii, these planets are larger by a factor of $\sim$1.2--2, depending on which star they orbit. In the case that they orbit the secondary star and it is a bound late-type companion, then the planets would likely still have radii in the super-Earth regime due to the smaller radius of the host. A less likely scenario is that they orbit the secondary and the star is actually a background star, in which case their radii are more uncertain. The mean stellar densities from the transit parameter estimates for both candidates are consistent with each other and with an M dwarf, further suggesting EPIC\,211428897 is the host. We conclude that further study is warranted, as this is likely a real system of small planets orbiting a late-type star, but we do not validate it because of the uncertainty about which star is the host.

The multi-candidate star EPIC\,211413752 has a companion detected in Gemini AO imaging 4.73\arcsec\, away and 5.9 magnitudes fainter (estimated in the \kepler bandpass, see G18). The star is close enough that it lies within the smallest aperture in our multi-aperture photometry analysis, so we cannot determine which star is the host from the light curve alone. However, if the candidates orbit the secondary star, their radii would be bigger by a factor of 230 due to dilution from the primary ($\gamma_\mathrm{sec}$ in \autoref{tab:ao-companions}), corresponding to $\sim$3.2 and $\sim$5.3 \rsun. As the radius of the primary star from Keck/HIRES is 0.77$\pm$0.03 \rsun, we can rule out the possibility for these signals to originate from the secondary star. Finally, the transit parameter estimates for both candidates yield mean stellar densities that are consistent with each other and with the density of the primary star, and inconsistent with a low density background giant. We conclude that EPIC\,211413752 is a valid host of a super-Earth and a sub-Neptune, and the planet radii listed in \autoref{tab:params} are accurate, as the dilution from the secondary star is negligible.

EPIC\,211491383 is a slightly evolved F star and the apparent host of a 1.6 \rearth\ planet candidate on a 4.1 day orbit, which has a low FPP of $7.7 \times 10^{-5}$. A nearby star $\sim$6 magnitudes fainter contributes $\sim$18\% of its flux to the optimal photometric aperture, and thus could conceivably be the host of the observed signal. Although the smallest aperture from our multi-aperture analysis excludes the flux of this nearby star, the resulting light curve is too noisy to draw any conclusions. If the transit signal were to come from the nearby faint star, the undiluted transit depth implies a radius ratio of $\sim40$\%. However, the best-fit transit model yields $T_{23} = 2.74$ and $T_{14} = 2.79$ hours, which is inconsistent with \Rp/\rstar$\approx$40\%, but consistent with the measured value of \Rp/\rstar$=1$\%. Furthermore, the stellar density implied by the transit fit is consistent with the target star. We conclude that the nearby star cannot be the host, and validate the planet orbiting EPIC\,211491383; the dilution from the nearby star is negligible, at less than 0.1\%.

EPIC\,211529065 is a multi-candidate system with a secondary star detected by {\it Gaia} just within the photometric aperture. The secondary star is $\sim$2.3 magnitudes fainter than the target star, and $\sim$80\% of its flux is within the aperture. The two planet candidates have orbital periods of 1.5 and 4.4 days, and assuming they orbit EPIC\,211529065, they have radii of approximately 1.4 and 3.0 \rearth. However, if they actually orbit the secondary star, these candidates would be $\sim$10 times bigger, taking into account dilution and assuming the secondary is the same size as that target. However, eclipses of the secondary by two $\sim$14 and $\sim$30 \rearth objects is {\it a priori} very unlikely. Furthermore, the {\it Gaia} DR2 parallax of the secondary implies it is a background giant, in which case the radii of the occulting objects would be larger, making this scenario even more unlikely. Finally, the smallest aperture in our multi-aperture analysis excludes more flux from the secondary than the optimal aperture, yet there is no apparent decrease in transit depth. Based on the above and the very low FPPs of the candidates, we conclude that EPIC\,211529065 is the true host of two validated small planets, and we note that their radii may be underestimated by up to $\sim$5\% due to dilution, but this is within the uncertainties.

EPIC\,212008766 hosts a single $\sim$2.2 \rearth\ planet candidate on a 14.1 day orbit. However, the optimal aperture selected by \ktwophot includes a nearby star $\sim$3.2 magnitudes fainter. Our multi-aperture analysis clearly shows that the signal originates from the primary star, as the small aperture excludes the flux of the secondary and there is no decrease in the transit signal. However, $\sim$90\% of the flux from the secondary is likely to be diluting the transit as measured from the optimal aperture light curve extraction. We validate the planet, but we note that the planet radius we report may be underestimated by $\sim$6\%, which is about the same size as the uncertainty.

EPIC\,212418133 is the apparent host of a low FPP ($9.0 \times 10^{-5}$) 3.0 \rearth planet on a 3.3 day orbit, but there is a star $\sim$6 magnitudes fainter $\sim$14\arcsec\ away, just outside the photometric aperture. Even though it contributes only $\sim$17\% of its flux, we cannot rule out the faint source as the host based on the computed value of the undiluted transit depth alone (i.e. maximum eclipse depth = 100\%). However, the smallest aperture in our multi-aperture analysis excludes essentially all of the flux from this source, but there is no apparent decrease in transit depth. We conclude that EPIC\,212418133 is the true host and we validate the planet.

The candidate 212435047.01 appears to be a $\sim$1.5 \rearth\ planet on a 1.1 day orbit, but $\sim$60\% of the flux of a nearby star is within the optimal \ktwophot aperture. Based on the observed transit depth and dilution, this source could potentially be the host, even though it is $\sim$7 magnitudes fainter. However, a more distant eclipsing binary on the same \kepler CCD column has a matching ephemeris (EPIC\,212409377), so this signal could also be caused by the ``column anomaly'' identified by \citet{2014AJ....147..119C}. We designate it as a candidate, but we note that it is most likely an instrumental false positive.

The candidate 212555594.01 was previously validated \citep[K2-192\,b][]{2018AJ....155..136M}, but a star $\sim$2 magnitudes fainter then EPIC\,212555594 $\sim$14\arcsec\ away contributes non-zero flux to the aperture such that it could potentially be the host. However, the signal can still be seen in the light curve from the smallest aperture, which excludes more of the flux from the neighbor than the optimal aperture. We conclude that the signal is indeed coming from EPIC\,212555594 and validate the planet.

The candidate 213951550.01 is almost certainly a false positive based on its FPP of 96\%, its transit depth of nearly 10\%, and its large radius of $\sim$23 \rearth. Furthermore, there is a nearby star $\sim$1.5 magnitudes brighter than the target contributing $\sim$19\% of its flux to the aperture. However, our multi-aperture analysis strongly suggests that the nearby star is not the source of the signal. In addition, there is significant out-of-transit stellar variability in phase with the transit signal suggestive of ellipsoidal variations. The target star appears to be an M dwarf with a radius of $\sim$0.5\rsun, so the system is most likely an eclipsing binary involving a second, lower-mass M dwarf.

We do not validate the candidate 220209578.01 because of its high FPP and large radius. The optimal aperture contains significant flux from a star $\sim$18\arcsec\ away and $\sim$2 magnitudes brighter, but our multi-aperture analysis shows that the signal does in fact come from EPIC\,220209578. However, the transit depth is diluted by a factor of $\sim$8, so the radius is in fact much larger than what we measure. This candidate is very likely a false positive, and the likelihoods in \autoref{tab:vespa-cand} suggest that it is a simple eclipsing binary scenario.

220448185.01 is listed in \autoref{tab:params} as a candidate USP with a FPP of 94\%. The optimal aperture contains significant flux from a fainter star $\sim$7.5\arcsec\ away ({\it Gaia} DR2 2564954125578601472), and our multi-aperture analysis indicates that this other star is actually the host of the transit signal, as the depth clearly decreases when flux from this neighbor is excluded. Preliminary inspection of the light curve suggests that the orbital period is actually half of the value reported by P18, which is likely the result of their decision to restrict their transit search to periods greater than 0.5 days. However, because the candidate is most likely an eclipsing binary, the transit-like features actually correspond to primary and secondary eclipses, in which case the orbital period reported by P18 is correct.

Nineteen of our candidates are previously confirmed planets that do not meet our validation criteria, although most have low FPPs that are more consistent with planetary than false positive scenarios: 211319617.01 \citep[K2-180\,b;][]{2018AJ....155..136M}, 211351816.01 \citep[K2-97\,b;][]{2016AJ....152..185G}, 211355342.01 \citep[K2-181\,b;][]{2018AJ....155..136M}, 211418729.01 \citep[K2-114\,b;][]{2017AJ....154..188S}, 211442297.01 \citep[K2-115\,b;][]{2017AJ....154..188S}, 211945201.01 \citep[EPIC\,211945201\,b][]{2018AJ....156....3C}, 211990866.01 \citep[K2-100\,b][]{2017AJ....153...64M}, 212110888.01 \citep[K2-34\,b;][]{2016ApJ...825...53H}, 212580872.01 \citep[K2-193\,b;][]{2018AJ....155..136M}, 212697709.01 \citep[WASP-157\,b;][]{2016PASP..128l4403M}, 212735333.01 \citep[K2-197\,b;][]{2018AJ....155..136M}, 212803289.01 \citep[K2-99\,b;][]{2017MNRAS.464.2708S}, 215969174.01 \citep[HATS-36\,b;][]{2018AJ....155..119B}, 216414930.01 \citep[HATS-11\,b;][]{2016AJ....152...88R}, 216468514.01 \citep[K2-107\,b;][]{2017AJ....153..130E}, 217671466.01 \citep[HATS-9\,b;][]{2015AJ....150...33B}, 218131080.01 \citep[HATS-12\,b;][]{2016AJ....152...88R}, 218916923.01 \citep[K2-139\,b;][]{2018MNRAS.475.1765B}, and 220504338.01 \citep[K2-113\,b;][]{2017MNRAS.471.4374E}.
Eleven of these candidates are not in our validated sample because they have radii larger than 10 \rearth (see \autoref{sec:large-radii}), and four have FPPs above our validation threshold of \minfpp. See \autoref{tab:params} for the parameters and FPPs of these systems. We note that candidates not validated because of their large radii were previously reported to have radii larger than 10 \rearth, with the exception of K2-97\,b, which was originally reported to have a radius of 14.7$\pm$1.2 \rearth \citep{2016AJ....152..185G,2017AJ....154..254G} and subsequently reported to have a radius of $8.04^{+1.43}_{-0.98}$ \rearth by \citet{2018AJ....155..136M}. We do not validate six of these candidates because of the presence of (bright) AO or {\it Gaia} DR2 sources within the photometric apertures, as described in \autoref{sec:companions}. We note that only three of these six candidates have been confirmed by radial velocity measurements (HATS-11\,b, HATS-12\,b, and K2-139\,b). The other three \citep[K2-180\,b; K2-193\,b; K2-197\,b;][]{2018AJ....155..136M} may warrant further observations to determine whether the detected signals originate from the primary or secondary stars; although \citet{2018AJ....155..136M} used light curves from a different photometric pipeline (and thus different apertures), the secondary stars are near and bright enough that they may contribute flux to even the smallest usable apertures.

One of our unvalidated candidates, 212572439.01, was designated a false positive by \citet{2017AJ....154..207D}, but we find that this disposition may be overly conservative. While we expect some of our unvalidated candidates to be false positives (especially those with high FPPs), 212572439.01 has a relatively low FPP of 0.7\%, and we did not validate it because of the bright secondary star contributing within the photometric aperture (see \autoref{fig:aper1}). The analysis of light curves from multiple \ktwo photometric pipelines (and thus different photometric apertures) by \citet{2017AJ....154..207D} yielded an inconsistent set of transit depths for this candidate, which they interpreted as being indicative of a blended EB scenario. However, even in the case that the signal is from the fainter secondary star, the candidate is still potentially in the planetary size regime, accounting for dilution from the primary and the {\it Gaia} DR2 stellar radius of the secondary. Indeed, our multi-aperture light curve analysis suggests that the signal may originate with the other star, and given the relatively similar radii of the two stars, the FPP is not likely to be significantly higher. EPIC\,212572439 and its neighbor thus warrant further observations to reveal the true nature of the signal.

On the other hand, the candidate 212773272.01 illustrates the importance of catalog queries and pixel-level analyses. Based on the light curve from the optimal aperture, this candidate's FPP is well below our validation threshold, but red flags were raised by both of these quality control checks. Examination of our {\it Gaia} DR2 query revealed that the signal could have originated from the brighter nearby star in the aperture, and the multi-aperture analysis showed hints of a radius dependence as well as a more pronounced ``V'' transit shape for the largest aperture. We therefore conclude that the FPP of from \vespa is invalid, and that 212773272.01 may be a false positive scenario similar to the blended eclipsing binaries reported by \citet{2017A&A...606A..75C}.

Another interesting case is that of 219388192.01. We did not validate this candidate because it had a measured radius above 10 \rearth (see \autoref{sec:large-radii}), and it also has several {\it Gaia} DR2 sources within the \ktwophot aperture that are bright enough to be the source of the observed transit signals (see \autoref{fig:aper2}). However, our multi-aperture analysis showed that the signals originate from the presumed host star, EPIC\,219388192. A search of the literature revealed that RV measurements have in fact revealed that this is a transiting brown dwarf \citep{2017AJ....153..131N}. This demonstrates the necessity for caution when statistically validating large planets, as they can be the same size as brown dwarfs or even low mass stars, as pointed out by \citet{2017ApJ...847L..18S}.

Our pipeline assumes a linear ephemeris, so the presence of uncorrected TTVs make the phase-folded transit more V-shaped, which affects our planet parameter estimates, as well as the likelihoods computed by \vespa. See \citet{2018AJ....155..127H} for an analysis that accounts for TTVs.

The candidate 212443457.01 is a likely false positive. P18 noted that this is a likely hierarchical triple system based on the appearance of the light curve. The middle of the three transits observed by \ktwo is likely to be a deep secondary eclipse, as it is visibly shallower than the other two transits.

\startlongtable
\begin{deluxetable*}{lcccccccc}
\tablecaption{\vespa likelihoods. Note: $a$ -- likelihood that the signal is due to a background eclipsing binary, at the measured period or twice that; $b$ -- likelihood that the signal is due to an eclipsing binary, at the measured period or twice that; $c$ -- likelihood that the signal is due to a hierarchical star system with an eclipsing component, at the measured period or twice that; $d$ -- likelihood that the signal is due to a planet. \label{tab:vespa-cand}}
\tablehead{ID &     L\_beb$^a$ &  L\_beb\_Px2$^a$ &          L\_eb$^b$ &      L\_eb\_Px2$^b$ &         L\_heb$^c$ &     L\_heb\_Px2$^c$ &     L\_pl$^d$ &     FPP}
\startdata
211319617.01     &                    $0$     &                    $0$     &    $2.1 \times 10^{-9}$     &   $5.4 \times 10^{-8}$     &   $5.8 \times 10^{-32}$     &  $4.0 \times 10^{-15}$     &   $6.3 \times 10^{-3}$     &   $8.9 \times 10^{-6}$  \\
211331236.01     &                    $0$     &                    $0$     &   $1.7 \times 10^{-11}$     &   $4.3 \times 10^{-9}$     &   $5.2 \times 10^{-50}$     &  $4.5 \times 10^{-37}$     &   $6.9 \times 10^{-1}$     &  $2.5 \times 10^{-10}$  \\
211331236.02     &   $4.7 \times 10^{-8}$     &   $8.6 \times 10^{-8}$     &   $5.8 \times 10^{-18}$     &   $1.3 \times 10^{-8}$     &   $9.1 \times 10^{-64}$     &  $3.9 \times 10^{-19}$     &   $8.7 \times 10^{-2}$     &   $6.7 \times 10^{-8}$  \\
211342524.01     &                    $0$     &                    $0$     &    $4.8 \times 10^{-2}$     &   $4.4 \times 10^{-4}$     &    $1.0 \times 10^{-3}$     &   $5.0 \times 10^{-5}$     &   $9.8 \times 10^{-5}$     &                    $1$  \\
211351816.01     &                    $0$     &                    $0$     &    $2.7 \times 10^{-4}$     &   $2.9 \times 10^{-6}$     &   $5.0 \times 10^{-16}$     &  $4.7 \times 10^{-11}$     &   $5.2 \times 10^{-4}$     &   $3.4 \times 10^{-1}$  \\
211355342.01     &                    $0$     &                    $0$     &    $2.6 \times 10^{-4}$     &   $2.9 \times 10^{-4}$     &   $2.4 \times 10^{-23}$     &  $3.5 \times 10^{-11}$     &   $1.8 \times 10^{-2}$     &   $3.0 \times 10^{-2}$  \\
211359660.01     &                    $0$     &                    $0$     &   $1.5 \times 10^{-16}$     &   $2.1 \times 10^{-8}$     &  $7.2 \times 10^{-215}$     &  $1.3 \times 10^{-28}$     &   $7.4 \times 10^{-1}$     &   $2.8 \times 10^{-8}$  \\
\enddata
\end{deluxetable*}

\begin{deluxetable}{lccc}
\tablecaption{Improvement in orbital period estimates from joint analysis of C5 and C16 light curves for a subset of candidates, where $\Delta_\mathrm{prec}$ is the factor by which the precision of the period estimate is improved by C5+C16, as compared to C5-only. \label{tab:c16}}
\tablehead{Candidate & $P_\mathrm{C5}$ & $P_\mathrm{C5+C16}$ & $\Delta_\mathrm{prec}$ \\
 & [days] & [days] & }
\startdata
211816003.01     &  $14.45353311^{+0.00132848}_{-0.00135966}$     &  $14.45354808^{+0.00004218}_{-0.00004271}$     &       32X  \\
212069861.01     &  $30.95152372^{+0.00362482}_{-0.00332025}$     &  $30.95530038^{+0.00021038}_{-0.00027763}$     &       14X  \\
212099230.01     &   $7.11233845^{+0.00028579}_{-0.00029388}$     &   $7.11241444^{+0.00001138}_{-0.00001103}$     &       26X  \\
211919004.01     &  $11.71633207^{+0.00117151}_{-0.00119811}$     &  $11.71954405^{+0.00002719}_{-0.00003048}$     &       41X  \\
212154564.01     &   $6.41376553^{+0.00030071}_{-0.00030252}$     &   $6.41366051^{+0.00001009}_{-0.00000971}$     &       30X  \\
212164470.02     &   $7.81025325^{+0.00084577}_{-0.00084378}$     &   $7.80919210^{+0.00002799}_{-0.00002502}$     &       32X  \\
212006344.01     &   $2.21928224^{+0.00012819}_{-0.00013011}$     &   $2.21929641^{+0.00000866}_{-0.00000886}$     &       15X  \\
211945201.01     &  $19.49221006^{+0.00066616}_{-0.00065325}$     &  $19.49215972^{+0.00001859}_{-0.00001840}$     &       36X  \\
211490999.01     &   $9.84398447^{+0.00057410}_{-0.00059092}$     &   $9.84401139^{+0.00006230}_{-0.00005837}$     &       10X  \\
212110888.01     &   $2.99563346^{+0.00000794}_{-0.00000790}$     &   $2.99563154^{+0.00000063}_{-0.00000063}$     &       13X  \\
211529065.01     &   $4.40002829^{+0.00016852}_{-0.00017814}$     &   $4.39981866^{+0.00000830}_{-0.00000870}$     &       20X  \\

\enddata
\end{deluxetable}

\end{document}